\documentclass[10pt,journal,compsoc]{IEEEtran}

\usepackage{times}

\usepackage{soul}
\usepackage{url}
\usepackage[hidelinks]{hyperref}
\usepackage[utf8]{inputenc}
\usepackage[small]{caption}
\usepackage{graphicx}
\usepackage{amsmath}
\usepackage{booktabs}
\usepackage{color}
\usepackage{xcolor}
 \usepackage{array}
\usepackage{multirow}
\usepackage{multicol}
\usepackage{amssymb}
\usepackage{diagbox}
\usepackage{float}
\usepackage{booktabs}
\usepackage{colortbl}
\usepackage{enumitem}
\usepackage{amsmath,amssymb,amsfonts}
\usepackage{algorithmic}
\usepackage{longtable}
\usepackage{threeparttable}
\newcommand{\tabincell}[2]{\begin{tabular}{@{}#1@{}}#2\end{tabular}}
\usepackage[linesnumbered,boxed,ruled,commentsnumbered]{algorithm2e} 

\let\olditem\item\renewcommand{\item}[1][black]{\color{#1}\olditem}

\newcommand{\revise}[1]{\textcolor{black}{#1}}
\newcommand{\revisetwo}[1]{\textcolor{black}{#1}}
\newcommand{\revisethree}[1]{\textcolor{black}{#1}}

\SetKwInOut{Input}{\textbf{Input}}
\SetKwInOut{Output}{\textbf{Output}}
\SetKw{Continue}{continue}
\SetKw{Break}{break}

\begin{document}
\title{Joint Admission Control and Resource Allocation of Virtual Network Embedding via Hierarchical Deep Reinforcement Learning}

\author{
    Tianfu~Wang,
    Li~Shen,~\IEEEmembership{}
    Qilin~Fan,~\IEEEmembership{}
    Tong~Xu,~\IEEEmembership{}
    Tongliang~Liu,~\IEEEmembership{}
    and~Hui~Xiong,~\IEEEmembership{Fellow,~IEEE}% <-this % stops a space
\IEEEcompsocitemizethanks{
\IEEEcompsocthanksitem{
This research was supported in part by the National Natural Science Foundation of China (Grant No.92370204) and the Science and Technology Planning Project of Guangdong Province (Grant No. 2023A0505050111).

\emph{(Corresponding author: Hui Xiong and Li Shen).}
Part of this work was done during Tianfu Wang’s internship at JD Explore Academy. }
\IEEEcompsocthanksitem Tainfu Wang and Tong Xu are with Anhui Province Key Laboratory of Big Data Analysis and Application, School of Computer Science, University of Science and Technology of China, Hefei 230027, China.
E-mail: tianfuwang@mail.ustc.edu.cn, tongxu@ustc.edu.cn.
\IEEEcompsocthanksitem Qilin Fan is with the School of Big Data and Software Engineering, Chongqing University, ChinaScience, Chongqing 401331, China.
E-mail: fanqilin@cqu.edu.cn.
\IEEEcompsocthanksitem Li Shen is with JD Explore Academy.
E-mail: mathshenli@gmail.com.
\IEEEcompsocthanksitem Tongliang Liu is with the UBTECH Sydney Artificial Intelligence Centre and the School of Information Technologies, Faculty of Engineering and Information Technologies, The University of Sydney, Darlington, NSW 2008, Australia. 
E-mail: tongliang.liu@sydney.edu.au.
\IEEEcompsocthanksitem Hui Xiong is with the Hong Kong University of Science and Technology (Guangzhou) Thrust of Artificial Intelligence Nansha, Guangzhou, 511400, Guangdong, China and the Hong Kong University of Science and Technology, Department of Computer Science and Engineering, Hong Kong SAR, China.
E-mail: xionghui@ust.hk.
}% <-this % stops an unwanted space
% \thanks{Submitted to IEEE TSC}
}

\markboth{IEEE Transactions on Services Computing,~Vol.~xx, No.~x, March~xxxx}%
{Shell \MakeLowercase{\textit{et al.}}: Bare Demo of IEEEtran.cls for Computer Society Journals}



\IEEEtitleabstractindextext{%

\begin{abstract}
As an essential resource management problem in network virtualization, virtual network embedding (VNE) aims to allocate the finite resources of physical network to sequentially arriving virtual network requests (VNRs) with different resource demands. Since this is an NP-hard combinatorial optimization problem, many efforts have been made to provide viable solutions. However, most existing approaches have either ignored the admission control of VNRs, which has a potential impact on long-term performances, or not fully exploited the temporal and topological features of the physical network and VNRs. In this paper, we propose a deep \textbf{H}ierarchical \textbf{R}einforcement \textbf{L}earning approach to learn a joint \textbf{A}dmission \textbf{C}ontrol and \textbf{R}esource \textbf{A}llocation policy for VNE, named HRL-ACRA. Specifically, the whole VNE process is decomposed into an upper-level policy for deciding whether to admit the arriving VNR or not and a lower-level policy for allocating resources of the physical network to meet the requirement of VNR through the HRL approach. Considering the proximal policy optimization as the basic training algorithm, we also adopt the average reward method to address the infinite horizon problem of the upper-level agent and design a customized multi-objective intrinsic reward to alleviate the sparse reward issue of the lower-level agent. Moreover, we develop a deep feature-aware graph neural network to capture the features of VNR and physical network and exploit a sequence-to-sequence model to generate embedding actions iteratively. Finally, extensive experiments are conducted in various settings, and show that HRL-ACRA outperforms state-of-the-art baselines in terms of both the acceptance ratio and long-term average revenue. Our code is available at \url{https://github.com/GeminiLight/hrl-acra}.
\end{abstract}

}

\maketitle

\IEEEdisplaynontitleabstractindextext

% \fi
\IEEEpeerreviewmaketitle

\IEEEraisesectionheading{\section{Introduction}\label{sec:introduction}}
\IEEEPARstart{A}{s} the demand for Quality of Service (QoS) has been continuously strengthened, research on efficient network resource management architectures has received massive attention from academia and industry.
Traditional network management architectures directly execute diverse network functions on different dedicated servers, confronted with the increasing pressure of network services.
Network virtualization has emerged as one of the promising approaches to overcome this problem, which is capable of decoupling the network services from their underlying hardware and empowering the programmability of services \cite{vne-nfv}.
By integrating advanced technologies such as software-defined networking (SDN) and network function virtualization (NFV), network virtualization has become an essential part of next-generation networks for its superior capabilities to improve network resource utilization and reduce network difficulty management significantly \cite{icc-2021-drl-sfcp}.

In the network virtualization framework, user network requests are constructed as graph-structured virtual network requests (VNRs), consisting of virtual network functions and virtual links, dynamically arriving at the physical network to acquire resources. Then, the Internet providers will attempt to allocate node and link resources of physical network for VNRs under various constraints.
To maximize the revenue of Internet providers, we need to decide which VNRs to be accepted and how resources to be allocated, constituting an NP-hard combinatorial optimization problem, called virtual network embedding (VNE) \cite{vne-nphard}.

For this fundamental network management problem, many research efforts have been devoted to improving the performance of VNE. Classical approaches are categorized as the exact \cite{vne-rd,vne-ilp,vne-tsc-multi-domain}, the heuristic \cite{vne-grc,vne-nrm,vne-tpds-dc} and the meta-heuristic \cite{vne-aco,vne-pso,vne-dc} ones.
For exact algorithms, they aim at finding the optimal embedding solutions yet result in high computational complexity.
Heuristic methods require rich expert knowledge and trial experience to design efficient heuristics. At the same time, meta-heuristic approaches are often non-deterministic and depend on extensive searches to discover high-quality solutions.

Recently, several learning-based approaches \cite{vne-neuro,vne-mcts,vne-a3c-gcn} have been proposed due to their excellent abilities to build better heuristics from data automatically. However, these works pose two challenges. First, they rely on manually-extracted features or merely exploit partial latent temporal or topological features of the physical network and VNRs. Insufficient representation capability will limit the accuracy of VNE decisions. Second, most existing approaches can only optimize resource allocation for the current incoming VNR, ignoring the admission control of VNRs which has a potential impact on long-term benefits. VNRs need to compete for the limited resources provided by the physical network. An embedded VNR will release the resource occupied until its service completes. \revisetwo{If a VNR is embedded with a low-quality solution, the network system will remain in a low resource utilization state until this VNR completes. As a result, more VNRs arriving later may be rejected due to scarce resources. Therefore, the uncertainty of upcoming VNRs presents a critical challenge in promoting the long-term benefits of Internet providers.}

To address the above-mentioned challenges, we propose a \textbf{H}ierarchical \textbf{R}einforcement \textbf{L}earning (RL) approach to learn a joint \textbf{A}dmission \textbf{C}ontrol and \textbf{R}esource \textbf{A}llocation policy for online VNE, named HRL-ACRA. Admission control and resource allocation are considered as upper-level and lower-level tasks in a hierarchical framework. We model both tasks as Markov decision processes (MDPs) and 
develop a hierarchical optimization framework based on the proximal policy optimization (PPO) \cite{drl-ppo} algorithm to train the corresponding policies. For each arriving VNR, an upper-level policy is responsible for deciding whether to admit it or not. The early rejection of inappropriate VNRs could reduce the unnecessary embedding process, thus improving resource utilization and accelerating the deployment speed. A lower-level policy is appointed to allocate resources of the physical network to meet the requirement of VNR.
\revise{Specially, the average revenue method \cite{drl-apo} is adopted to address the infinite horizon problem of the upper-level agent, and a multi-objective intrinsic reward is customized to alleviate the sparse reward issue of the lower-level agent.}
To fully use topological information and temporal relationship, we also design some deep neural network modules for sufficient feature extraction.
Thanks to these carefully designed modules, HRL-ACRA achieves state-of-the-art (SOTA) performance with respect to several criteria. 

To the end, our contributions are summarized as four-fold:

\begin{itemize}
%   \item We investigate the admission control-aware virtual network embedding problem, considering multi-type resources. We cast admission control and resource allocation as upper-level and lower-level tasks in a hierarchical framework, model them as MDPs, and utilize hierarchical RL to learn a joint policy.
    \item We propose a hierarchical RL method for the admission control-aware online VNE problem. We cast admission control and resource allocation as upper-level and lower-level tasks in a hierarchical framework, model them as MDPs, and learn a joint policy. The upper-level agent takes the long-term benefits into consideration, while the lower-level agent pays close attention to the short-term profit.
    \item To address the infinite horizon problem caused by the continuous interactions between the upper-level agent and incoming VNRs, we adopt the average revenue method to enhance the upper-level agent to treat the current and future revenue more equally.
    We also customize a multi-objective intrinsic reward with multiple local indicators to alleviate the sparse reward issue, encouraging the lower-level agent to perform exploration.    
    \item We design a deep feature-aware graph neural network (GNN) with link feature awareness to make full use of topological features of the physical network and VNRs. Exploiting the topological information and temporal dependence simultaneously, we employ a sequence-to-sequence (Seq2seq) model based on this GNN and gated recurrent unit (GRU) to generate embedding actions iteratively.
    \item To simulate various real network situations, we conduct simulation experiments by adjusting the environment parameters. Extensive experimental results demonstrate that the proposed approach outperforms other SOTA algorithms on acceptance ratio and long-term average revenue.

\end{itemize}

The remainder of this paper is organized as follows. The related work is presented in Section \ref{related-work}. 
In Section \ref{preliminaries}, we provided the formulation of VNE problem and the basic concepts of GNN and RL. 
The details of our proposed algorithms are described in Section \ref{methodology}.
In Section \ref{experiments}, the experimental results and their analysis are given.
Finally, we conclude this paper in Section \ref{conclusion}.

\section{Related Work}
\label{related-work}

\begin{table*}[t]
\caption{The Summary of Related Works}
\centering
\begin{tabular}{cccclll}
\hline
\multicolumn{1}{c}{Work} & \multicolumn{1}{c}{AC} & Online & Method Category &  Methodology & Feature Extraction & Exploited features\\ \hline
\cite{vne-rd} & $\times$ & $\checkmark$ & exact  & MILP & /  & / \\
\cite{vne-ilp} & $\times$ & $\checkmark$ & exact & ILP & / & / \\
\cite{vne-tsc-green} & $\times$ & $\checkmark$ & exact & ILP and MILP & /  & / \\
\hline
% \cite{vne-noderank} & $\times$ & $\checkmark$ & heuristic & Node ranking & Random walk  & Only physical network \\ 
\cite{vne-grc} & $\checkmark$ & $\checkmark$ & Heuristic & Node ranking & Global resource control  & Physical network and virtual network \\ 
\cite{vne-nrm} & $\times$ & $\checkmark$ & Heuristic & Node ranking & Node resource management & Physical network and virtual network \\
\revise{\cite{vne-tpds-dc}} & \revise{$\times$} & \revise{$\checkmark$} & \revise{Heuristic} & \revise{Priority of location} & \tabincell{l}{\revise{Evaluation on node and path}} & \revise{Physical network and virtual network} \\

\cite{vne-tsc-delay-vnf} & $\times$ & $\checkmark$ & Heuristic  & Recursion and backtracking & / & / \\  

\hline

\cite{vne-aco} & $\times$ & $\checkmark$ & Meta-heuristic  & Ant colony optimization  & / & / \\
\cite{vne-pso} & $\times$ & $\checkmark$ & Meta-heuristic  & Particle swarm optimization  & / & / \\ 
\cite{vne-dc} & $\times$ & $\checkmark$ & Meta-heuristic & Overlapping decomposition & / & / \\ 
\hline
\cite{vne-ac-rnn} & $\checkmark$ & $\checkmark$ & SL & Classification & RNN & \tabincell{l}{Physical network and virtual network} \\
\hline
\cite{vne-neuro} & $\times$ & $\checkmark$ & uSL  & Subgraph exaction & Hopfield network & \tabincell{l}{Only physical network} \\ 
\cite{vne-gae-bfs} & $\times$ & $\checkmark$ & uSL  & Clustering  & GAE  & \tabincell{l}{Only physical network} \\ \hline
\cite{vne-mcts} & $\times$ & $\checkmark$ & RL  & MCTS & / & \tabincell{l}{Physical network and virtual network} \\ 
\cite{vne-mlp} & $\times$ & $\checkmark$ & RL  & REINFORCE & MLP & \tabincell{l}{Physical network and current virtual node} \\ 
\cite{vne-td} & $\times$ & $\checkmark$ & RL & Temporal-Difference & MLP & \tabincell{l}{Physical network and virtual network} \\ 
\cite{vne-cnn-double-layer} & $\times$ & $\checkmark$ & RL  & REINFORCE & CNN  & \tabincell{l}{Only physical network} \\
\cite{vne-pg-rnn} & $\times$ & $\checkmark$ & RL  & REINFORCE & RNN  & \tabincell{l}{Only Physical Network} \\
\cite{vne-a3c-gcn} & $\times$ & $\checkmark$ & RL  & A3C & GCN & \tabincell{l}{Physical network and current virtual node} \\  
\cite{vne-hrl} & $\checkmark$ & $\times$ & RL & Hierarchical RL with DQN & GCN & \tabincell{l}{Physical network and current virtual node} \\  
Ours & $\checkmark$ & $\checkmark$ & RL  &  \tabincell{l}{Hierarchical RL with PPO \\ Average reward method} & \tabincell{l}{Seq2Seq Model \\ Customized GNN} & \tabincell{l}{Physical Network, virtual network and \\ VNR's global attributes} \\  
\hline
\end{tabular}
\label{table-related-work}
\end{table*}

Since the VNE problem is critical but challenging in network virtualization, studies for tackling it have received tremendous attention.
% \cite{vne-survey}
In this section, we summarize the related works in two categories, the traditional and learning-based methods, and their brief information is concluded in Table \ref{table-related-work}.

\subsection{Traditional Methods} 

\revise{We categorize the traditional methods into three categories, the exact, heuristic, and meta-heuristic algorithms.}

% exact
For the exact algorithms, the VNE problem can be typically formulated as mathematical programming.
Chowdhury \textit{et al}. \cite{vne-rd} transformed VNE as a mixed-integer linear programming (MILP) by physical network augmentation and constraint relaxation, coordinating node, and link Mapping.
Shahriar \textit{et al}. \cite{vne-ilp} formulated VNE as an integer linear programming (ILP) problem, taking spare capacity location and request survivability into consideration. 
Diallo \textit{et al}. \cite{vne-tsc-multi-domain} developed an ILP-based method for VNR splitting and a MILP-based approach for resource mapping to solve the multi-domain VNE problem.
However, these algorithms cannot handle online deployment scenarios for their high computational complexity.

% heuristic
Instead of achieving optimal solutions, a wealth of  heuristics-based algorithms are proposed to find feasible solutions quickly. Among them, node ranking is the primary strategy. 
\revise{Gong \textit{et al}. \cite{vne-grc} applied global resource capacity to evaluate the importance of nodes of the physical network and virtual network based on a random walk model, and then used the greedy matching strategy and the breadth-first search algorithm to conduct the node mapping and link mapping, respectively. A simple admission control strategy based on revenue-to-cost also was proposed to improve the long-term benefits. Still, it only focused on the embedding solution of the current VNR without considering future VNRs.
Similarly, Zhang \textit{et al}. \cite{vne-nrm} adopted a capacity-based metric to sort nodes with multiple resource types.}

Even though these algorithms are capable of generating feasible solutions quickly, they incur high blocking rates.
Some schemes are also proposed to further improve the solution quality of VNE problems by handling failure situations.
Yang \textit{et al.} \cite{vne-tsc-delay-vnf} studied the NP-hardness of the delay-sensitive and availability-aware VNE problem, and introduced a recursive heuristic method using the restricted shortest path algorithm method to solve it.
Fan \textit{et al}. \cite{vne-tpds-dc} designed a priority of location VNE algorithm, following node proximity sensing and path comprehensive evaluation.
Lin \textit{et al}. \cite{vne-tsc-energy} designed several heuristic algorithms for energy-aware VNE problem based on auxiliary graph building methods.
Nevertheless, designing heuristics relies on rich expert knowledge and trial experience and only accommodates a few specific scenarios.

Moreover, many meta-heuristic algorithms have also been adopted to address the VNE problem and obtain near-optimal solutions.
Fajjari \textit{et al}. \cite{vne-aco} presented an ant colony optimization algorithm for VNE problems that iteratively searches the solution spaces of node mapping.
Su \textit{et al}. \cite{vne-pso} designed an energy-aware VNE algorithm based on the particle swarm optimization technique, regarding the node mapping solution as the position of particles.
To solve large-scale VNE problems, Song \textit{et al}. \cite{vne-dc} proposed a divide-and-conquer evolutionary algorithm that utilizes the overlapping decomposition method to group the critical elements with tight connections to many other nodes into multiple sub-components. 
However, meta-heuristic algorithms are often non-deterministic and depend on extensive searches to discover high-quality solutions.

\subsection{Learning-based Methods} 

Recently, machine learning has proved to be a promising direction for solving combinatorial optimization problems. 
According to the nature of the available learning signal, machine learning is usually categorized into three major categories: supervised learning (SL), unsupervised learning (uSL), and reinforcement learning (RL).
Several works investigate machine learning applications to the VNE problem, and we introduce some of them following the three categories mentioned above.

% supervised learning & % unsupervised learning
Supervised learning has achieved great success in image recognition, text classification, and other scenarios. Treating the admission control mechanism of VNE as a binary classification problem, Andreas \textit{et al}. \cite{vne-ac-rnn} applied this idea to improve the existing VNE algorithm's performance by utilizing a Recurrent Neural Network (RNN) to judge whether to admit or reject arriving VNRs, but only considered the rejection of infeasible VNRs to avoid wasting time.
Different from supervised learning relying on labeled samples, unsupervised learning is capable of discovering underlying structures in unlabeled data.
To reduce the large search space, Blenk \textit{et al}. \cite{vne-neuro} applied the Hopfield neural network to extract subgraphs fed to other existing VNE algorithms, resulting in faster and more resource-efficient embeddings.
Habibi \textit{et al}. \cite{vne-gae-bfs} adopted the graph autoencoder (GAE) to cluster physical nodes and then randomly sample one node as the initial center to execute the breadth-first search (BFS) method in each cluster.

% reinforcement learning
The embedding process of VNE can be regarded as a series of decisions, and some studies tackled this problem by reinforcement learning (RL). 
Modeling the node selection process as a Markov decision process (MDP), Haeri \textit{et al}. \cite{vne-mcts} utilized the Monte Carlo tree search (MCTS) to improve the revenue-to-cost ratio.
Wang \textit{et al}. \cite{vne-td} trained a multi-layer perceptron (MLP) model with temporal-difference learning to approximate the value function of VNE states.
Using a convolutional neural network (CNN) to extract the attributes of physical nodes and training it with REINFORCE algorithm, Li \textit{et al}. \cite{vne-cnn-double-layer} proposed a double-layer RL-based framework for VNE to learn the node mapping and link mapping process jointly.
Yao \textit{et al}. \cite{vne-pg-rnn} designed a sequence-to-sequence model based on a recurrent neural network (RNN) to exploit the historical information of previous decisions, and train with the REINFORCE algorithm.
\revisethree{But these traditional neural networks often exhibit limited representational capabilities when dealing with non-Euclidean data, requiring manual topological feature extraction like closeness and betweenness from graph theory. In contrast, graph neural networks (GNNs) excel at automatically mining deep features from graph data, generating richer representations \cite{gnn-survey}.}
Yan \textit{et al}. \cite{vne-a3c-gcn} presented an RL-based method with a graph convolutional network (GCN) \cite{gnn-gcn} to extract state features from the physical network, but did not explicitly exploit edge features.
Cheng \textit{et al}. \cite{vne-hrl} proposed a hierarchical RL control framework to select the VNR from batch candidates stored in the time window, where the GCN is used for the rough feature exploitation. However, their subagent for admission control only works for offline settings, e.g., within the time window batch, and is not unqualified to tackle the online VNE where future VNRs are not known in advance.
In conclusion, existing RL-based works only concentrate on the resource allocation of arriving VNRs, or exploit partial latent temporal or topological features of the physical network and VNRs.

Compared with existing learning-based works, HRL-ACRA learns a joint admission control and resource allocation policy based on a hierarchical framework, capable of perceiving the long-term benefits and short-term interests. The well-designed neural network model also enables extracting sufficient temporal relationships and topological structures.

\section{Preliminaries}
\label{preliminaries}

\subsection{Admission control-aware Online VNE}

\subsubsection{System Description}
\begin{figure}[t]
    \setlength{\belowcaptionskip}{-0.5cm}
	\centering
	\includegraphics[width=.45\textwidth]{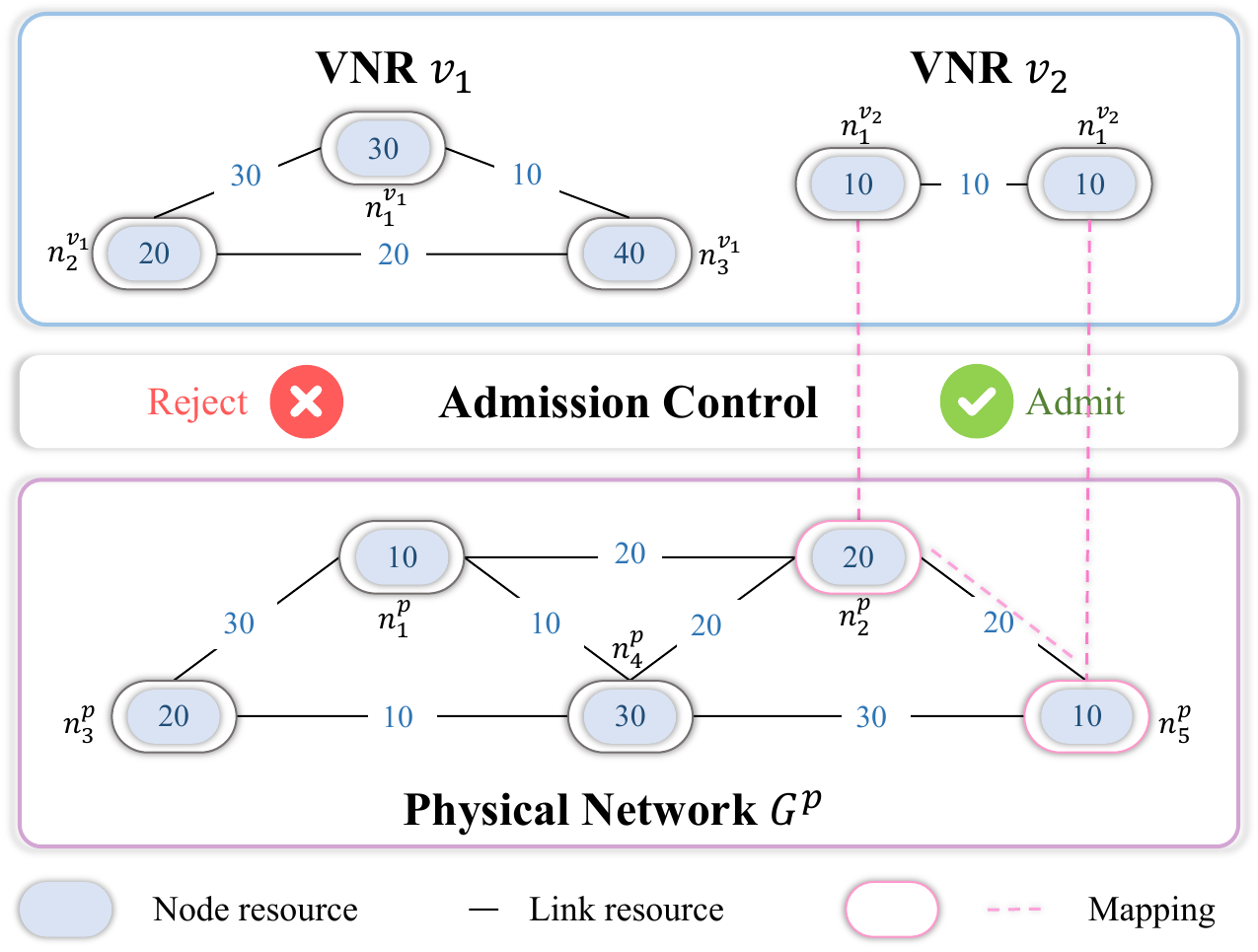} 
	\caption{\revisetwo{An example of admission control-aware online VNE problem. For VNR $v_1$, one of its virtual nodes, $n^{v_1}_4$, requires 40 units of node resources, while no physical node in the physical network $G^p$ has enough available node resources to satisfy such demand. Due to constraint violations caused by insufficient resources, there is no feasible solution for $v_1$. Therefore, admission control  early rejected $v_1$ to skip the resource allocation process. In the case of VNR $v_2$, all the resource demands of its virtual nodes and links can be accommodated by the physical network, satisfying all the constraints. As a result, $v_2$ can be admitted and successfully embedded. A feasible solution is shown with pink lines, where $v_2$'s nodes $n_1^{v_1}$ and $n_1^{v_2}$ are embedded into physical nodes $n^p_2$ and $n^p_5$, respectively, and $v_2$'s link $(n_1^{v_1}, n_1^{v_2})$ is embedded into a physical path $[(n^p_2, n^p_5)]$. Hereafter, these mapped physical nodes (links) will update their available resources by subtracting the resource demand of corresponding virtual nodes (links).}
 }
	\label{fig-example}
\end{figure}

Figure \ref{fig-example} illustrates an example of the admission control-aware VNE Problem. 
In the practical network system, user services are virtualized as the virtual network requests (VNRs) set $V$, dynamically requesting resources provided by the physical network. 

\begin{itemize}[leftmargin=*]
    \item  \textit{Physical Network} is defined as a weighted undirected graph $G^p = (N^p, L^p)$, where $N^p$ and $L^p$ respectively denote a set of physical nodes and a collection of physical links. The remaining node resources capacity of each physical node $n^p \in N^p$ is represented by $C_{n^p}$, while the remaining link resources capacity of each physical link $l^p \in L^p$ is characterized as $B_{l^p}$.
    \item \textit{Virtual Network Request} is defined as a tuple $v = \langle G^v, A^v \rangle$, where $G^v$ denotes a virtual network and $ A^v$ represents its global attributes. Here, $A^v$ denotes the global attributes, e.g., lifetime. We describe the virtual network as a weighted undirected graph $G^v = (N^v, L^v)$, where $N^v$ refers to a set of virtual nodes, and $L^v$ is a collection of virtual links. Similarly, each virtual node $n^v \in N^v$ specifies node resources demand $C_{n^v}$, while each virtual link $l^v \in L^v$ indicates link resources demand $B_{l^v}$.
\end{itemize}

% Admission control-aware VNE
In standard online VNE workflow, the system will always attempt to allocate resources of physical network resources to each arriving VNR. The VNR will only be accepted if all virtual nodes and virtual links are mapped to physical nodes and physical paths, respectively, under the condition that the constraints are satisfied. 
Once a VNR is successfully embedded, the resources will be occupied until the service completes. Since embedded VNRs cannot be discarded until their lifetime expires, each incoming VNR needs to compete for limited resources provided by the physical network with the currently serving VNRs. 
Therefore, the uncertainty of upcoming VNRs results in a critical challenge to promoting the long-term benefits of Internet providers.

\begin{figure*}[t]
    \setlength{\belowcaptionskip}{-0.5cm}
	\centering
	\includegraphics[width=.94\textwidth]{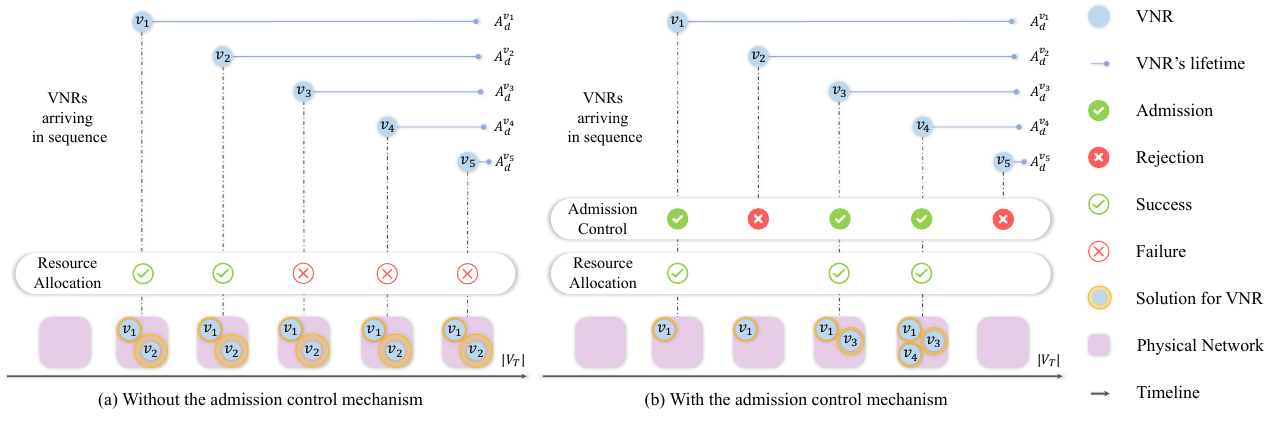}
	\caption{\revisetwo{Comparison of systems with and without admission control mechanism. (a) Without the admission control mechanism, all arriving VNRs will be attempted to execute resource allocation directly. After admitting and embedding VNR $v_1$, this system embeds VNR $v_2$ with a low-quality solution and the majority of physical resources are conquered by $v_1$, contributing to the rejection of $v_3$, $v_4$, and $v_5$ for the scarcity of physical resources. The resulting long-term average and acceptance ratio of this system are $LA\_Rev_{a} = \frac{\sum_{i \in \{1, 2\}} (w_a + w_b * A^{v_1}_d)  Rev(G^{v_i})}{|V^{T}|}$ and $AC\_Ratio_{a} = \frac{2}{5}$, respectively. (b) With the admission control mechanism, dynamically arriving VNRs are selectively admitted and enter the resource allocation stage. Admission control proactively rejects  VNR $v_2$ owing to the difficulty to find a high-quality solution, avoiding long-term low resource utilization. Then, sequentially arrived $v_3$ and $v_4$ are admitted and embedded successfully. For VNR $v_5$, the system early rejects it due to the lack of available physical resources, skipping the unnecessary resource allocation process. The resulting long-term average and acceptance ratio of this system are $LA\_Rev_{b} = \frac{\sum_{i \in \{1, 3, 4\}} (w_a + w_b * A^{v_i}_d) Rev(G^{v_i})}{|V^{T}|}$ and $AC\_Ratio_{b} = \frac{3}{5}$, respectively.
    Using admission control makes the acceptance of $v_3$ and $v_4$ possible by actively rejecting $v_2$, improving resource utilization. Compared with the absence of admission control mechanism, it achieves better performance on both the acceptance ratio and long-term average revenue. By early rejecting VNRs having no feasible solution, it also saves inference time and enhances the real-time decision.}}
	\label{fig-ac-motivation}
\end{figure*}

Admission control (AC) is a proactive scheduling mechanism that has a potential impact on long-term benefits. 
It decides whether to admit or reject the incoming VNR, according to the VNR's requirements and the current situation of physical network. 
Considering the resource availability of the current physical network and the future VNR demand, the system can adaptively reject some VNRs to reserve resources for subsequent VNRs to improve long-term benefits, including acceptance ratio and long-term average revenue.
\revisetwo{For the unadmitted VNRs, they can further refine their orchestration \cite{vne-orchestration} to enhance the probability of finding a high-quality feasible solution and request again.}
\revise{Additionally, early rejecting of VNRs in which no feasible solution exists and skipping the resource allocation phase contributes to speeding up the decision-making.} 
\revise{Figure \ref{fig-ac-motivation} shows a comparison of without and with the admission control mechanism. 
}
Here, we emphasize that the status of admission control has long been ignored by most existing works. In this work, we try to utilize the hierarchical RL method to solve the VNE problem with more practical and challenging settings by jointly considering admission control and resource allocation. The detailed description of the admission control-aware VNE procedure is summarized in Algorithm \ref{algo-ac-vne}.

\begin{algorithm}[ht]
	\caption{Admission Control-aware VNE Procedure}
	\label{algo-ac-vne} 
% 	\begin{algorithmic}[1]
    \Input{\revise{A VNE solver $S$;} The arriving VNR $v = \langle G^v, A^v \rangle$, where $G^v= (N^v, L^v)$; \\
    The current physical network $G^p = (N^p, L^p)$;}
    \Output{The embedding result for the VNR $v$;}
    \BlankLine
    /**Admission Control**/; \\
    \eIf{$v$ is Admitted}{
        \revise{$S$ executes the Resource Allocation process;}\\
    }
    {
        \Return False;\\
    }
    /**Resource Allocation**/; \\
    \For{Virtual node $n^v_i \in N^v$, $i = 1, 2, \cdots$} {
        \revise{$S$ finds a physical node $n^p_i$ for virtual node $n^v_i$;} \\
        \revise{$S$ attempts to place $n^v_i$ onto $n^p_j$;}\\
        \eIf{$n^v_i$ is successfully placed}{
            \For{Virtual node $n^v_j \in N^v$, $i = 1, 2, \cdots$}{
            \If{$(n^v_i, n^v_j) \notin L^v$}{
                \Continue;\\
            }
            \revise{$S$ finds a physical path $p^p_{i,j}$ for $(n^v_i, n^v_j)$;} \\
            \revise{$S$ attempts to route $(n^v_i, n^v_j)$ onto $p^p_{i,j}$;}\\
            \eIf{$(n^v_i, n^v_j)$ is successfully routed}{
                \Continue;\\
            }{
                \revise{$S$ undoes all the previous actions;} \\
                \Return False;\\
            }
        }
        }{
            \revise{$S$ undoes all the previous actions;} \\
            \Return False;\\
        }
    }
    \Return True;\\
\end{algorithm}

\subsubsection{Problem Formulation} 

Embedding a VNR into the physical network can be defined as a mapping process from ${G}^v$ to a subgraph of physical network ${G}^{p'}$ with various constraints. 
\begin{equation}
f: {G}^v \rightarrow {G}^{p'}.
\label{eq-gm}
\end{equation}
It is composed of node mapping and link mapping, where massive discrete variables need to be decided.

\begin{itemize}[leftmargin=*]
    \item  \textit{Node Mapping} is defined to assign all virtual nodes $n^v \in N^v$ to feasible physical nodes $n^p \in N^p$, constrained by:
    \begin{equation}
    \sum_{n^p \in N^p} \phi^{n^v}_{n^p} = 1, \forall{n^v \in N^v}, \label{eq-nm-1}
    \end{equation}
    \begin{equation}
    \sum_{n^v \in N^v} \phi^{n^v}_{n^p} \leq 1, \forall{n^p \in N^p}, \label{eq-nm-2}
    \end{equation}
    \begin{equation}
    \revisethree{\phi^{n^v}_{n^p} C_{n^v} \leq C_{n^p}, \forall{n^v \in N^v}, \forall{n^p \in N^p}}, \label{eq-nm-3}
    \end{equation}
    where $\phi^{n^v}_{n^p}$ is a binary variable that is set 1 when $n^v$ is embedded in $n^p$. $C_{n^v}$ and $C_{n^p}$ denote node resource requirement and remaining capacity. Eq. \eqref{eq-nm-1} and Eq. \eqref{eq-nm-2} are related to the relationship between every virtual node $n^v$ and physical node $n^p$. Each virtual node $n_v \in G^v$ of the same virtual network must be placed on different network nodes. Eq. \eqref{eq-nm-3} refers to the node resource constraint, i.e., the available resources of each physical node must exceed the resource request of the virtual node carried by it.
    \item \revise{ \textit{Link Mapping} is defined to embed each virtual links $l^v \in L^v$ into one loop-free physical path $p^p_{l_v} \in P^p$, constrained by:
    \begin{equation}
    B_{l^v} \leq B_{l^p}, \forall{l^p \in p^p_{l_v}} \label{eq-lm}
    \end{equation}
    where $P^p$ is defined as a set of all loop-free physical paths in the physical network. $p^p_{l_v}$ is one element of $P^p$ that connects two physical nodes that accommodate the source and destination nodes of the virtual link $l^v$.
    $B_{l^v}$ and $B_{l^p}$ denote the bandwidth resource requirement and remaining capacity. 
    Eq. \eqref{eq-lm} means the available bandwidth of each physical link in the physical path must exceed the bandwidth request of the virtual link. The pink lines in Figure \ref{fig-example} depict an example of the mapping process of one virtual link.
    }
\end{itemize}

\revisetwo{If node mapping or link mapping fails due to constraint violations, it indicates that a feasible solution for the VNR has not been found, which means the VNR cannot be embedded.}

\subsubsection{Performance Evaluation} \label{performance-evaluation}
The most important performance evaluation metrics are acceptance ratio ($AC\_Ratio$), long-term average revenue ($LA\_Rev$), and running time \cite{vne-survey}, which are defined as:
\begin{itemize}[leftmargin=*]

\item  \textit{Acceptance Ratio} measures the QoS with the number of the accepted VNRs, which is defined as:
\begin{equation}
{AC\_Ratio} = \lim_{T\to \infty}\frac{\sum_{v\in V_T}\mathbb{I}(v)}{|V_T|},
\end{equation}
where $\mathbb{I}(v)$ is the indicator function that returns 1 if the VNR $v$ is accepted and 0 otherwise.

\item  \textit{Long-Term Average Revenue} is a direct indicator to reflect the revenue of Internet providers, which is defined as:
\begin{equation}
 LA\_Rev =\lim _{T \rightarrow \infty}  \frac{\sum_{v\in V_T} \mathbb{I}(v) (w_a + w_b * A^v_d)  \operatorname{Rev}\left({G}^{v}\right)}{|V_T|},
\end{equation}

where $V_T=\{v|0<t_v<T\}$ is the set of VNRs arriving before time instance $T$. $A^v_d$ donates the lifetime of VNR $v$. 
$w_a$ and $w_b$ are the weight of starting price and the weight of the service time charge, respectively. 
Internet providers can make various pricing strategies by deciding the value of $w_a$ and $w_b$ to control the two components of price.
$\text{Rev}({G}^{v})$ is the basic revenue of a VNR calculated by the total amount of resource requests of its nodes and links, computed by
\begin{equation}
   \text{Rev}({G}^{v}) = \sum_{n^v \in N^v} C_{n^v} + \sum_{l^v \in L^v} B_{l^v}
\end{equation}

\item  \textit{Running Time} describes the time consumption of the VNE algorithm to process VNRs. 
In practice, a VNE algorithm needs to face online and dynamic scenarios, which requires a high real-time guarantee. 
Consequently, it is necessary that the used VNE algorithm should balance the running time and performance, arranging the arriving VNR as soon as possible to meet the real-time requirements.

\end{itemize}

\revisetwo{In this work, we jointly optimize the long-term average revenue and acceptance ratio, while considering running time.}

\subsection{Graph Neural Network}
Graph neural networks (GNN) that can operate on non-European data have become a hot topic in the field of deep learning recently, which enables deep learning technology to be competent for more complex tasks. Based on the message propagation mechanism between nodes, GNN models the relationships and dependencies between nodes to extract the deep-level information in the graph-structured data \cite{gnn-survey}. A variety of GNNs can be divided into several categories, one of which is the spatial-based graph convolutional neural network. By designing an aggregation function and adopting a message propagation mechanism, the spatial-based graph convolutional neural network updates the central node's representation by aggregating its neighbor nodes' representation.

Graph attention network (GAT) \cite{gnn-gat} is a typically spatial-based graph convolutional neural that integrates the self-attention mechanism into the propagation step. To obtain the new representation of one node, GAT computes adaptive attention coefficients between two nodes to aggregate representations of neighbor nodes. Formally, the graph convolutional operation of GAT $\ell$-th layer is defined by:
\begin{equation}
    h_{i}^{(\ell)} = \sigma(\sum_{j \in {N}(i) \cup\{i\}} \alpha_{i,j} W h_j^{(\ell-1)})
    \label{eq-gat-operation}
\end{equation}
where $h_{i}^{(\ell)}$ and $h_i^{(\ell-1)}$ are the node representation of node $i$ in the $\ell$-th and $\ell-1$-th GNN layer, respectively. $\sigma$ is an activation function and $W$ is a learnable weight to linear project the node input node representation. The coefficient $\alpha_{i,j}$ between node
i and node j is computed as:
\begin{equation}
    \alpha_{i, j}=\frac{\exp \left(\sigma\left(W_a\left[ h_{i}^{(\ell)} \| h_{j}^{(\ell) }\right]\right)\right)}{\sum_{k \in {N}(i) \cup\{i\}} \exp \left(\sigma\left(W_a \left[h_{i}^{(\ell)} \| h_{k}^{(\ell)} \right]\right)\right)}.
    \label{eq-gat-coef}
\end{equation}
Here, $\|$ is the concatenation operation. $N(i)$ is the set of node $i$'s neighbor nodes and $W_a$ is a trainable attention weight vector. 

\revisetwo{
Graph pooling networks are usually utilized to obtain the global representation of one whole graph for conducting graph-level tasks. Graph attention pooling (GAP) is one of the powerful graph pooling networks that considers both node features and graph topology based on self-attention \cite{gnn-gap}. GAP aggregates all node representations $h$ by doing a weighted summation, where a context is employed to compute the attention coefficients. The context is defined as the linear transformation of all node representations' mean pooling: 
\begin{equation}
    c = \sigma \left(\left(\frac{1}{\left|N\right|} \sum_{i=1}^{\left|N\right|} h_{i}\right) W\right),
\label{eq-context}
\end{equation}
where $\left|N\right|$ denotes the number of nodes and $W$ is a learnable weight matrix. Using this context $c$, the attention coefficient of node $i$ is given by :
\begin{equation}
    a_{i}=h_i^{Tr} c.
\end{equation}
Here, $Tr$ represents the transpose operation. Finally, the graph-level representation is computed as the weighted sum of all node representations:
\begin{equation}
g = \sum_{i}^{\left|N\right|} a_{i} h_{i}.
\end{equation}
}

\subsection{Reinforcement Learning}

Reinforcement Learning (RL) is an experience-driven learning framework widely used to solve sequential decision-making problems, where
an agent continuously interacts with an environment to collect experiences for learning.
\revise{In RL, Markov decision process (MDP) serves as a general mathematical model. It can be defined as a tuple $(\mathcal{S}, \mathcal{A}, P, R, \gamma)$, where $\mathcal{S}$ denotes the state space of environment, $\mathcal{A}$ represents the action space of agent. $P: \mathcal{S} \times \mathcal{A} \rightarrow S$ indicates the transition function from the current state to the next state in response to the selected action. $R: \mathcal{S} \times \mathcal{A} \times \mathcal{S} \rightarrow \mathbb{R}$ is the immediate reward function to evaluate the decision quality at the current state, mapping from states and actions into real numbers.}
$\gamma \in [0, 1]$ is a scalar discount factor to regulate the agent's attention to short-term rewards and long-term returns.
Specifically, the agent selects an action $a_t$ based on the observed state $s_t$ from the environment at each decision timestep $t$. After the agent executes the action $a_t$, the environment will transition to a new state $s_{t+1}$ and feedback a reward $r_t$ to the agent.
The objective of the agent is to find an optimal policy $\pi^{*}$, a mapping function from states into actions, to maximize the expected return:
\begin{equation}
\pi^{*}=\underset{\pi}{\operatorname{argmax}} \mathbb{E}\left[ G_t \right],
\end{equation}
where $G_t = \sum\limits_{}^{t} \gamma^{t} r_t$ denotes cumulative discounted reward.

Policy gradient is a kind of mainstream reinforcement learning algorithm, which computes an estimator of the policy gradient for gradient ascent algorithm to optimize the policy.
The optimization objective of the most classic gradient estimator is
\begin{equation}
L(\theta)=\hat{\mathbb{E}}_{t}\left[\log \pi_{\theta}\left(a_{t} \mid s_{t}\right) \hat{A}_{t}\right],
\end{equation}
where $\pi_{\theta}$ is a policy parameterized by $\theta$ and $\hat{A}_{t}$ is an advantage estimator at timestep $t$. 
The simplest advantage estimator uses the average return over several episodes as the baseline.
More practically, a learnable value function can assist in the advantage estimator, which estimates the expected return of a given state:
\begin{equation}
V_{\pi_{\theta}}(s_t) = \hat{\mathbb{E}}_{t}\left[ G_t | s_t \right].
\end{equation}
These approaches in such form are unified under the actor-critic architecture where the actor (policy) learns a policy to mask decisions and the critic (value function) estimates the expected return of the state. Benefit from the critic working as a baseline, the advantage estimator can be computed by:
\begin{equation}
\hat{A}_{t} = r_t + V_{\pi_{\theta}} (s_{t+1}) - V_{\pi_{\theta}}(s_t).
\label{eq-rl-adv}
\end{equation}
By doing this, the bias estimated by the value function is introduced to trade-off variance reduction policy gradients.

\section{Methodology}
\label{methodology}

\begin{figure}[ht] 
    \setlength{\belowcaptionskip}{-0.5cm}
	\centering
	\includegraphics[width=.45\textwidth]{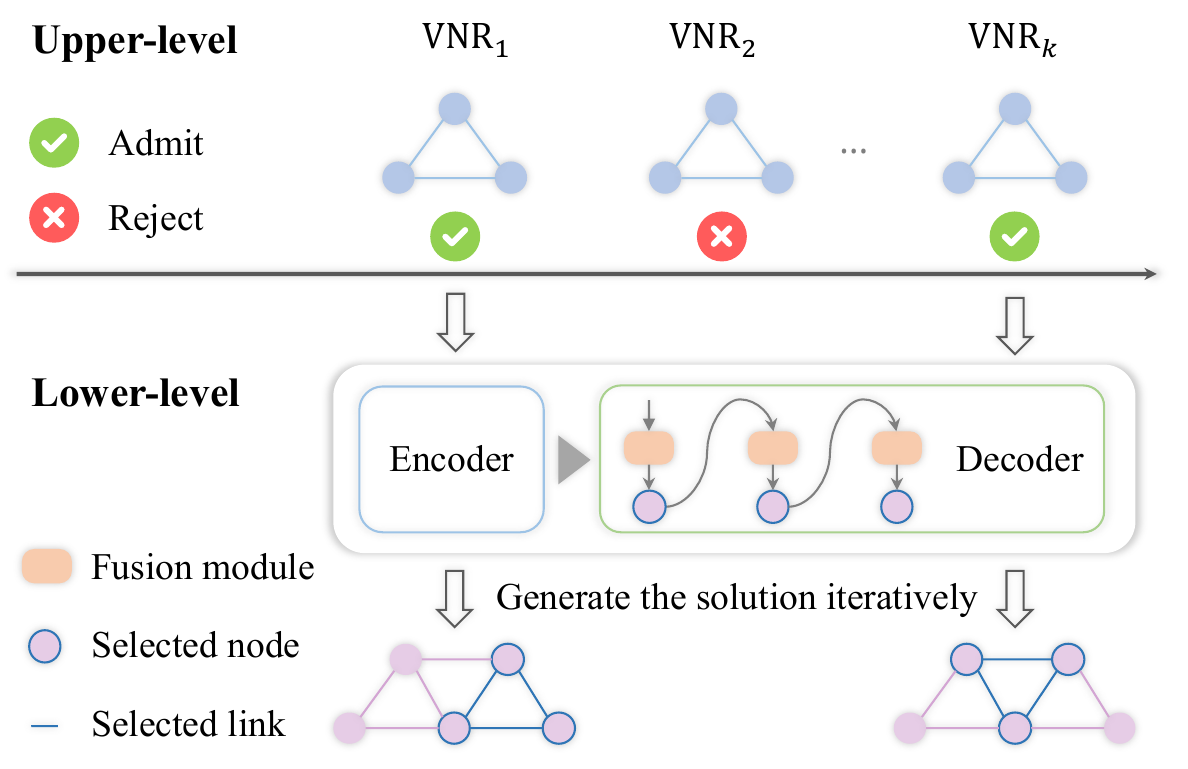} 
	\caption{Overall framework of our proposed hierarchical method for admission control-aware VNE.}
	\label{fig-framework}
\end{figure}

To solve this challenging combinatorial optimization problem of VNE, we propose a \textbf{H}ierarchical \textbf{R}einforcement \textbf{L}earning based approach to learn a joint \textbf{A}dmission \textbf{C}ontrol and \textbf{R}esource \textbf{A}llocation policy of VNE, named HRL-ACRA. As illustrated in Figure \ref{fig-framework}, we cast the admission control and resource allocation as upper-level and lower-level tasks, respectively. 
The upper-level agent has a foresight capability, whose goal is to optimize the long-term benefits, including the acceptance ratio and long-term revenue. 
According to the current situation of physical network and the request information of incoming VNR, it decides whether to admit the arriving VNR.
The lower-level agent focuses on the resource allocation for admitted VNRs to generate high-quality solutions. 
To reduce the action space, we utilize a sequence-to-sequence (seq2eq) model to construct the solution iteratively rather than in one step, i.e., the embedding actions of each virtual node are generated sequentially. Considering the state of VNRs is consistent, this seq2seq model consists of a static encoder and a dynamic decoder. The encoder extracts the decision order and features of nodes and links as node representations. The decoder selects a physical node to accommodate the current virtual node iteratively, based on the information aggregated by a fusion module at each timestep. Two agents are trained by the proximal policy optimization (PPO) method. 

\begin{figure*}[t] 
    \setlength{\belowcaptionskip}{-0.cm}
	\centering
	\includegraphics[width=.88\textwidth]{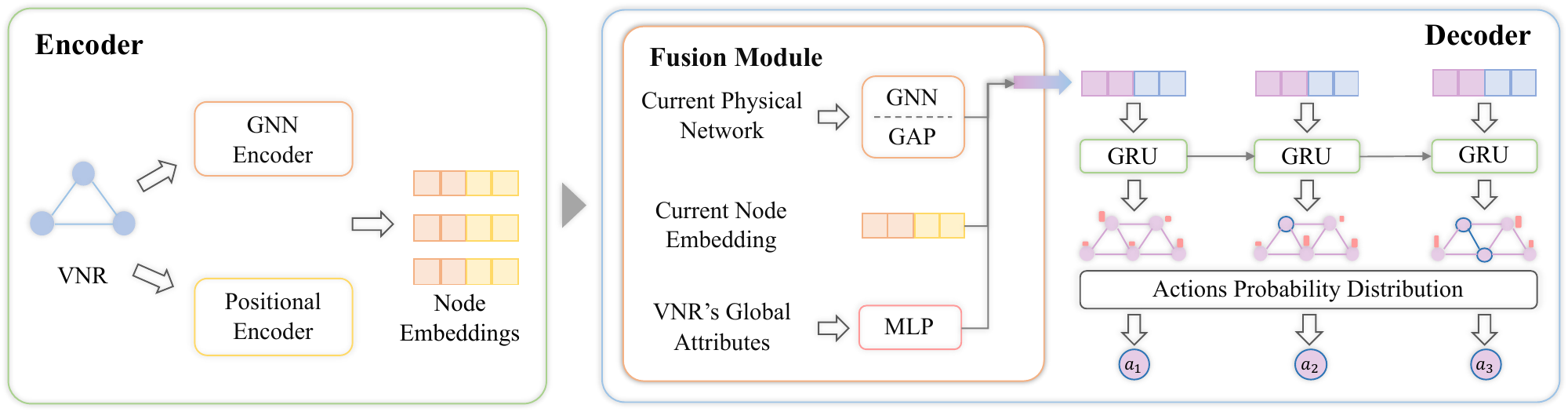} 
	\caption{Overview of the policy network of the lower-level agent. i) Encoder: The embeddings of each virtual node are composed of the feature embedding extracted by GNN Encoder and position embedding generated by the positional encoder. ii) Decoder: At each timestep $t$, the GRU-based decoder iteratively generates the embedded actions for each virtual node by aggregating the current embedding of the virtual node, the situation of physical network, and VNR's global attributes with the fusion module.}
	\label{fig-lower-agent}
\end{figure*}

\subsection{Upper-level Agent}
The upper-level agent is responsible for deciding on admission or rejection of arriving VNRs, which optimizes the acceptance ratio and the long-term revenue of Internet providers.
\subsubsection{MDP Definition} \revise{We model whether to admit the arriving VNR as an MDP, which contains three key elements: state, action, and reward.}
\begin{itemize}[leftmargin=*]
    \item \textbf{State}. The upper-level state consists of the request information of incoming VNR and the current situation of the physical network, shown in Table \ref{state}. Specifically, the request information of incoming VNR includes node resource demands and link resource requirements, topology structure, and lifetime. And the current situation of the physical network is composed of maximum and available resources of all nodes and links, and topology structure. 
    \item \revise{\textbf{Action}. The upper-level action $\hat{a} \in \hat{\mathcal{A}} = \{0, 1\}$ is a binary variable that indicates whether to admit the incoming VNR.}
    \item \revise{\textbf{Reward}. The extrinsic reward for the upper-level agent is designed to guide to optimize the long-term average revenue, also considering the revenue-to-cost ratio and avoidance of failed embedding to enable assessment of more situations, defined as:
    \begin{equation}
    \hat{r}= 
    \begin{cases}
    \frac{{\text{Rev}({G}^v)}}{\text{Cost}({G}^v)} \text{Rev}({G}^v),  & \text{if ${G}^v$ is admitted and embedded}, \\
    -{w_1}, & \text{if ${G}^v$ is admitted but not embedded},\\
    0, & \text{otherwise}.
    \end{cases}
    \end{equation}
    Here, $ \frac{{\text{Rev}({G}^v)}}{\text{Cost}({G}^v)}$ denotes the revenue-to-cost ratio. Similar to the $\text{Rev}(G^v)$, $\text{Cost}(G^v) =  \sum_{n^v \in N^v} C_{n^v} + \sum_{l_v \in L_v}\sum_{l_p \in p^p_{l_v}} B_{l^v}$ is computed by the total amount of resource consumption for embedding $G^v$. A higher revenue-to-cost ratio means fewer resources are consumed to embed the current VNR, saving more resources for future VNRs. 
    When an arriving VNR is admitted and then successfully embedded, we return a positive reward to encourage the agent to satisfy as many requests as possible. Considering different embedding solutions for the same VNR and the physical network, a better revenue-to-cost ratio can contribute to greater reward. For admitted but unsuccessfully embedded VNRs, we feedback a small negative reward $-{w_1}$ to prevent the agent from admitting infeasible VNRs, which contributes to omitting the resource allocation phase and improving the decision speed.}
\end{itemize}

\begin{table*}[t]
\caption{\revise{The Upper-level and Lower-level States}}

\centering
\begin{tabular}{cll}
\arrayrulecolor{black}
\hline
\multicolumn{1}{c}{} & \multicolumn{1}{c}{\revise{The Upper-level State}} & \multicolumn{1}{c}{\revise{The Lower-level State}} \\ \hline
\multirow{5}{*}{\revise{Physical Network}} 
 & \revise{Remaining resources of each physical node} & \revise{Remaining resources of each physical node} \\
 & \revise{Remaining resources of each physical link} & \revise{Remaining resources of each physical link} \\
  & \revise{Indexes of physical links} & \revise{Indexes of physical links} \\
 &  & \revise{Maximum resources of each physical node} \\
 &  & \revise{Maximum resources of each physical link} \\
 &  & \revise{Selection flag of physical nodes} \\
 &  & \revise{Neighbor flag of physical nodes} \\
 \hline
\multirow{3}{*}{\revise{Virtual Network}} 
 & \revise{Resource demands of each virtual node} & \revise{Resource demands of each virtual node} \\
 & \revise{Resource demands of each virtual link} & \revise{Resource demands of each virtual link} \\
 & \revise{Indexes of virtual links} & \revise{Indexes of virtual links} \\ \hline
\multirow{1}{*}{\revise{VNR's Global Attributes}} & \revise{Lifetime of the VNR} & \revise{Remaining number of virtual nodes to be embedded}
%  & Total number of virtual nodes & 1 
 \\ \hline
\end{tabular}
\label{state}
\end{table*}

\subsubsection{Policy Network}
\revisetwo{Since both the physical network and VNRs are graph-structured, GNN is a suitable and powerful feature extractor that is employed in our framework. GCN \cite{gnn-gcn} and GAT \cite{gnn-gat} are the two best representative GNN models.
GAT allows for adaptive weights for each neighbor node when aggregating the node features, while GCN assigns the fixed weight to all neighbor nodes. This attention mechanism enables GAT to adaptively model the importance of different neighborhood nodes and capture more fine-grained information about the neighborhood. Therefore, GAT is more powerful and flexible than GCN, especially in our tasks where fine-grained information about the graph structure is important.}

Unlike traditional graph tasks, link features are significant for VNE to perceive bandwidth resources. To engage bandwidth awareness with our deep model, we extend the GAT to blend link features into nodes during the propagation process.
% which usually cannot be extracted by existing graph neural networks (GNN).
% To aggregate node features and links features of physical network and VNRs, 
\revise{Besides, the shortest paths determined during the link mapping stage may consist of multiple hops. As such, it is necessary to build a deep GNN to learn a comprehensive node representation by leveraging information from multi-hop neighbors. However, the use of many layers in the construction of a GNN can lead to the problem of over-smoothing. To mitigate this issue, we deepen the GNN by using initial residual connections \cite{net-res} and identity mapping \cite{gnn-gcnii}.}

GAT is capable of extracting node features and topology information from non-European data. Still, it is not directly applicable to the VNE problem where link features play an important role. Consequently, we extend GAT to blend link features into node embedding, bringing awareness of bandwidth resources. Based on Eq. \eqref{eq-gat-coef}, introducing link features into the calculation of coefficients, the attention coefficient between node $i$ and $j$ is computed by:
\begin{equation*}
\alpha_{i, j}=\frac{\exp \left(\sigma\left(W_a\left[ h_{i}^{(\ell)} \left\| h_{j}^{(\ell)} \right\|  h_{i, j}\right]\right)\right)}{\sum_{k \in {N}(i) \cup\{i\}} \exp \left(\sigma\left(W_a \left[h_{i}^{(\ell)} \left\| h_{k}^{(\ell)} \right\| h_{i, k}\right]\right)\right)}.
\end{equation*}
where $h_{i, j} = \text{MLP}(B_{e_{i,j}})$ is the edge representation of link $e_{i,j}$, obtained by pass its link features $B_{e_{i,j}}$ into an MLP.

Moreover, deepening the number of GNN layers has been shown to be more effective in mining information from the graph, but it also confronts the challenge of over-smoothing. Therefore, we customize an attribute-aware GNN using the initial residual connection and identity mapping to alleviate this problem. Formally, the operation of this GNN $\ell$-th layer is defined by:
\begin{align}
h_{i}^{(\ell)} = \sigma\Big[\sum_{j \in {N}(i)  \cup\{i\}} ((1-\alpha) \alpha_{i,j} h_j^{(\ell-1)} + \alpha h_j^{0)} ) \nonumber\\ ((1-\beta) I_n + \beta W) \Big].
\end{align}
Here, $\alpha$ and $\beta$ are two hyperparameters denoting the strength of the initial residual connection and the identity mapping, respectively. ${h^0_j}$ is the initial node representation of node $j$. 
An identity mapping ${I}_n = f(H^{(\ell)})$ is added to the $\ell$-th weight matrix $W^{(\ell)}$.

The direct use of node representations is computation-intensive, we adopt graph attention pooling (GAP) to obtain the graph-level representation based on the attention mechanism \cite{net-attention}. 
Specifically, when a VNR $v$ arrives in the system, the physical network $G^p$ and the virtual network $G^v$ of this VNR are encoded into node representations as $\hat{h}^{p}$ and $\hat{h}^v$ respectively by this GNN:
\begin{equation}
\hat{h}^{p} = \text{GNN}\left( G^{p} \right), \hat{h}^{v} = \text{GNN}\left( G^{v} \right).
\end{equation}

We then adopt graph attention pooling (GAP) to obtain the graph-level representations of physical and virtual network. 
Meanwhile, VNR's global attributes ${A}^{v}$ are perceivable by MLP:
\begin{equation}
\hat{g}^{p} = \text{GAP}\left(\hat{h}^{p} \right), \hat{g}^{v} = \text{GAP}\left(\hat{h}^{v} \right), \hat{g}^{r} = \text{MLP}\left( {A}^v \right).
\end{equation}

We fuse them with an MLP and then pass the result and last hidden state $\hat{h}$ to a GRU to receive the current hidden state $\hat{h}^{\prime}$:
\begin{equation}
\hat{h}^{\prime} = \text{GRU}\left( \text{MLP}\left( \hat{g}^{p}, \hat{g}^{v}, \hat{g}^{r}\right), \hat{h} \right).
\end{equation}

The final upper-level action probabilities $\hat{\pi} $ is generated by:

\begin{equation}
\hat{\pi} = \text{Softmax} \left( \text{MLP} \left( \hat{h}^{\prime} \right) \right).
\end{equation}

\subsection{Lower-level Agent}

\subsubsection{MDP Definition}
\revisetwo{When a VNR is admitted by the upper-level agent, the lower-level agent will attempt to find an embedding solution to allocate resources for it. Similarly, we model the resource allocation process of VNE as an MDP, which is defined as follows.}
\begin{itemize}[leftmargin=*]
    \item \textbf{State}. \revisetwo{Compared to the upper-level state, the lower-level state needs more detailed information on the resource allocation situation for the same VNR. This is due to the state changes that occur after a virtual node is successfully embedded.} Specifically, the lower-level state is composed of the current situation of the physical network and virtual network, and VNR's global attributes, which are similar to the upper-level state. Notably, we design two binary flag features to reflect the status of physical nodes at each timestep, including
    (1) Selection flag of physical nodes indicates whether the physical nodes were selected. The flag value is 1 if the physical node was already selected; otherwise, 0; (2) Neighbor flag of physical nodes represents whether the neighbor of the current virtual node is placed in the physical node. If true, the value is 1; otherwise, 0. The detail of the lower-level state is shown in Table \ref{state}.
    \item \revise{\textbf{Action}. The lower-level action space ${\mathcal{\mathcal{A}}}$ is a subset of physical nodes, where we utilize a mask vector to dynamically remove nodes with insufficient resources or that have been selected before. An action $a \in \mathcal{\mathcal{A}}$ is selected to accommodate the current virtual node. Then, we utilize the  available graph construction-based shortest path method \cite{vne-grc} to execute the link mapping.}
    \item \textbf{Reward}. There is no guiding signal for the lower-level agent until the VNR is completely embedded or rejected, resulting in a sparse reward issue. To alleviate this issue, we develop a multi-objective intrinsic reward with multiple local indicators to guide the agent to perform efficient exploration, which is defined as:
    \begin{equation}
    r = 
    \begin{cases}
        \delta + \frac{\text{Rev}({G}^v)}{\text{Cost}({G}^v)}, & \text{if} \ G^v \ \text{is embedded successfully}, \\
         \delta, & \text{if} \ a_t \ \text{is successful}, \\
        -\frac{1}{|N^v|} & \text{otherwise},
    \end{cases}
    \end{equation}
    where $\delta = \frac{1}{|N^v|}\left( \frac{\text{Rev}_t({G}^v)}{\text{Cost}_t({G}^v)} + w_2 \psi (a_t)\right)$ is a temporal signal to estimate the quality of current partial solution at each timestep $t$, considering both newly revenue-to-cost ratio and load balancing.
     $\frac{\text{Rev}_t({G}^v)}{\text{Cost}_t({G}^v)}$ is the newly revenue-to-cost ratio computed by the revenue divided by the cost at timestep $t$. $\psi (a_t)$ is defined as the resource load balancing of physical nodes $a_t$, calculated by the ratio of the remaining resource to the maximum resource of $a_t$.
     $w_2$ denotes the weight of node load balancing. 
     A high revenue-to-cost ratio and resource load balancing can reserve more resources for subsequent VNRs. Once the $G^v$ is accepted, i.e., $t = |N^v| \text{ and } a_t \text{ is successful}$, we additionally return the complete revenue-to-cost ratio to award the agent for finding a feasible embedding solution.
\end{itemize}

\subsubsection{Policy Network}
As shown in Figure \ref{fig-lower-agent}, a seq2seq model, containing a static encoder and a dynamic decoder, is employed to extract the static and dynamic features respectively to seek a trade-off between running time and performance.

\begin{itemize}[leftmargin=*]
    \item \textbf{Encoder}. For the fixed state of VNR, we adopt a static GNN encoder to learn the feature embedding of each virtual node. Additionally, the decision order of each virtual node is usually preset, which is generally neglected by existing learning-based methods and incapable of being handled by GNN. Therefore, a position encoder \cite{net-transformer} is also used to generate the position embedding for each virtual node $h^v_p$, which is concatenated ($\text{CONCAT}$) with the virtual node's GNN embedding $h^v_f$ to form the VNR's node embedding $h^v$ as follow:
\begin{gather}
h^v_f = \text{GNN} \left( g^v \right), h^v_p = \text{PE} \left( g^v \right), \\
h^v = \text{CONCAT} \left( h^v_f, h^v_p \right). 
\end{gather}

 \item \textbf{Decoder}. 
At each timestep $t$, we encode the state of the physical network ${G}^{p}_t$ with GNN to extract node embeddings and use a GAP layer to congregate them to obtain the graph embedding. Simultaneously, the global attributes of VNR are sent into an MLP as follows:
\begin{equation}
\! g^{p}_t =\! \text{GAP}\left( \text{GNN}\left( {G}^{p}_t \right) \right), g^r =\! \text{MLP}\left( {A}^{v} \right).
\end{equation}
The fusion embedding at timestep $t$ is renewed by integrating the current virtual node embedding $h^v_t$, the VNR's global attributes representation, and the physical network's graph embedding. And then, the fusion embedding is input into a GRU, pulling in the last hidden state $h_{t-1}$:
\begin{equation}
h_t = \text{GRU}\left( \text{MLP}\left( g^{p}_t, h^v_t, g^{r}\right), h_{t-1} \right).
\end{equation}
Finally, we obtain the lower-level action probability distribution:
\begin{equation}
\pi_t = \text{Softmax}(\text{MLP}(h_t)).
\end{equation}
\end{itemize}

\subsection{Training Method}

\begin{algorithm}[ht]
	\caption{The training process of lower-level agent}
	\label{algo-pretrain-lower} 
% 	\begin{algorithmic}[1]
    \Input{The initial parameters of the lower policy $\theta$;}
    \Output{The optimized parameters of the lower policy $\theta$;\\} 
    \BlankLine
    \For{iteration$ = 1, 2, \cdots$ } {
        \For{$v_i, i = 1, 2, \cdots$} {
            Get the initial state $s_0$ from environment;\\
            $s_t \leftarrow s_0$;\\
            Encode the state of VNR with encoder;\\
            \For{Virtual node $n^v_j \in N^v$, $j = 1, 2, \cdots$} {
                Generates the action probability distribution $\pi_t$ with decoder;\\
                Sample an action $a_t$ based on $\pi_t$;\\
                Attempt to place $n^v_j$ onto $a_t$;\\
                \eIf{$n^v_j$ is successfully placed} {
                    Execute the link mapping;\\
                    
                    \If{Link mapping is unsuccessful} {
                        Undo all the previous actions;\\
                    }
                }
                {
                    Undo all the previous actions;
                }
                Receive the intrinsic reward $r_t$;\\
                Get the next state $s_{t+1}$;\\
                Collect experience \{$s_t, a_t, r_t, s_{t+1}$\};\\
                \If{Update}{
                    Optimize the parameters of policy $\theta$;\\
                }
                \If{Node placing or link mapping fails}{
                    \Break
                }
                $s_t \leftarrow s_{t+1}$;\\
            }
        }
    }
\end{algorithm}

\begin{algorithm}[ht]
	\caption{The training process of upper-level agent}
	\label{algo-joint} 
% 	\begin{algorithmic}[1]
    \Input{The initial parameters of the upper policy $\hat{\theta}$;\\
    }
    \Output{The optimized parameters of the upper policy $\hat{\theta}$;\\} 
    \BlankLine
    \For{iteration$ = 1, 2, \cdots$ } {
        Get the upper-level state $\hat{s}_0$; \\
        $\hat{s} \leftarrow \hat{s}_0$;\\
        \For{$v_i, i = 1, 2, \cdots$} {
            Upper-level agent generates action probability distribution $\hat{\pi}$; \\
            Upper-level agent samples action $\hat{a}$ based on  $\hat{\pi}$; \\
            \If{$\hat{a} = $  admit}{
                Lower-level agent attempt to allocate resources for $\text{VNR}_i$.
            }
            Upper-level agent receives the extrinsic reward $\hat{r}$;\\
            Upper-level agent gets the next state $\hat{s}^{\prime}$;\\
            Upper-level agent collects experience \{$\hat{s}, \hat{a}, \hat{r}, \hat{s}^{\prime}$\};\\
            $\hat{s} \leftarrow \hat{s}^{\prime}$;\\
            \If{Update}{
                Update the parameters of upper-level policy $\hat{\theta}$;\\
            }
        }
    }
\end{algorithm}

\begin{algorithm}[h]
	\caption{Inference with HRL-ACRA}
 
	\label{algo-inference} 
% 	\begin{algorithmic}[1]
    \Input{A series of VNRs $V$ arriving at the physical network sequentially;}
    \Output{Embedding results of each VNR $\Phi$;}
    \BlankLine
    \For{$v_i, i = 1, 2, \cdots$} {
        Upper-level agent decides whether to admit it;\\
        \eIf{admitted}{
            Lower-level agent  generates multiple solutions with beam search; \\
            % Encoder of lower-level agent encodes the VNR; \\
            % Decoder of lower-level agent; \\
            \eIf{exist feasible solutions}{
                Select the best solution to allocate resources; \\
                $\Phi \leftarrow \Phi \cup \{ \text{True} \}$;
            }
            {
                $\Phi \leftarrow \Phi \cup \{ \text{False} \}$;
            }
            
        }
        {
            $\Phi \leftarrow \Phi \cup \{ \text{False} \}$;
        }
    }
\end{algorithm}

To enhance the stability and efficiency of training, the proximal policy optimization (PPO) \cite{drl-ppo} is adopted to train two agents.
PPO is an improved policy gradient algorithm based on actor-critic architecture,  whose objective function is following:
\begin{equation}
L(\theta)=\min \left(r_{\theta} \cdot \hat{A}, \operatorname{clip}\left(r_{\theta}, 1-\epsilon, 1+\epsilon\right) \cdot \hat{A}\right),
\end{equation}
where $r_{\theta} = \frac{\pi_{\theta} (a_t | s_t)}{\pi_{\theta_{old}} (a_t | s_t)}$ denotes the probability ratio defined by the ratio of the current policy with parameter $\theta$ to the pre-update policy with parameter $\theta_{old}$, $\hat{A}$ represent an estimator of the advantage function computed by the discounted accumulative reward minus the prediction value of critic network, and $\epsilon$ indicates a hyperparameter clipping the probability ratio. 

The upper-level agent continuously tackles arriving VNRs in the practical scenario, whose interaction with the environment can be modeled as an infinite-horizon MDP.
The RL methods using discounted rewards are difficult to adapt to the infinite horizon MDP because they cannot treat the current and future revenue equally.
Therefore, we utilize the average reward method \cite{drl-apo} to modify the advantage estimator Eq. \eqref{eq-rl-adv} to allow the upper-level agent to learn policies that maximize average long-term returns:
\begin{equation}
\hat{A}_{t} = r_t - \rho + V_{\pi_{\theta}} (s_{t+1}) - V_{\pi_{\theta}}(s_t),
\end{equation}
where $\rho = \frac{1}{T} \sum_{t=1}^T r_t$ is the average reward for $T$ timesteps.

\revisethree{
Initially, we train the lower-level agent to learn a great resource allocation policy in the default experimental setting, where this agent tackles a diverse range of VNRs and endows its policy with excellent generalizability. As a result, this well-trained policy consistently produces high-quality solutions across various scenarios \cite{vne-a3c-gcn}, and we freeze it in the subsequent process.
Then, our focus then shifts towards training the upper-level agent, to learn adaptive admission control strategies tailored to different experimental settings. This process ensures both training efficiency and adaptability of HRL-ACRA.} 
We also apply the advantage normalization and entropy regularization \cite{drl-entropy} methods to improve the efficiency of training exploration. In the training phase, two agents sample actions according to generated policies. 
\revise{The detailed training scheme of two agents is summarized in Algorithms \ref{algo-pretrain-lower} and \ref{algo-joint}. Specially, both agents interact with their environment to gather experience. Each agent generates an action probability distribution based on current state and samples an action following this distribution. The environment transits to new states and feedbacks rewards to the agent.
At intervals of specified timesteps, they utilize this accumulated experience to optimize their parameters. 
}
\revisethree{Algorithm \ref{algo-inference} describes the inference workflow of HRL-ACRA for tackling a series of sequentially arriving VNRs. For each incoming VNR, the upper-level agent first decides whether to admit it. If admitted, the lower-level agent will parallelly generate multiple solutions with the beam search strategy \cite{acl-2017-beam-search} and select the best feasible solution to embed it, which contributes more sufficient exploration of solution space.}

\begin{table}
\caption{Parameter Settings}
\centering
\begin{tabular}{clc}
\hline
Group & \multicolumn{1}{c}{Parameter} & Value \\ 
 \hline
\multirow{8}{*}{Neural Network} 
 & Number of GRU layers & 1 \\
 & Hidden dimension of GRU & 128 \\
 & Number of MLP layers & 3 \\
 & Number of GNN layers & 5 \\
 & Embedding dimension of GNN & 128 \\
 & Strength of the initial residual connection & 0.2 \\
 & Strength of the identity mapping & 0.2 \\ 
 \hline
\multirow{3}{*}{Model Training}
 & Learning rate of actor & 0.0010 \\
 & Learning rate of critic & 0.0005 \\
 & Batch size & 256 \\
 & Discount factor of reward $\gamma$ & 0.99 \\
 \hline
\multirow{3}{*}{Reward Weights}
%  & Weight of revenue-to-cost ratio $w_1$ & 0.1 \\
 & Weight of negative reward $w_1$  & 0.1 \\
 & Weight of node load balancing $w_2$  & 0.01 \\
 \hline
\end{tabular}
\label{table-params}
\end{table}

\section{Experiments}
\label{experiments}
In this section, we conduct extensive experiments to demonstrate the effectiveness of HRL-ACRA algorithms, by summarizing the main implementation details and result analysis. Due to the limitation of page length, we place the baseline descriptions, ablation studies, and hyperparameter sensitivity analysis in Appendix.

\subsection{Experimental Setup}
\label{experimental-setup}

\subsubsection{Implementation Details}
Following the settings of most previous works \cite{vne-grc},\cite{vne-a3c-gcn},\cite{vne-mlp}, we implement a typical VNE simulator to evaluate the performance of multiple algorithms under the same platform. The topology of the physical network is generated following a Waxman random graph with the parameters $\alpha = 0.5$ and $\beta = 0.2$, representing a medium-sized network system with 100 nodes and about 500 links. The central processing unit (CPU) and bandwidth are considered as node and link resources, respectively.
The node and link resources of physical network satisfy the uniform distribution from 50 to 100 units.
\revisetwo{In one simulation of the training and testing phase, we randomly generate 1000 VNRs to arrive at the system successively following a Poisson process with an average arrival rate of 4 per 100 time units.}
Each VNR has an exponentially distributed lifetime with an average of 1000 time units. By default, every VNR has [2-10] nodes following the uniform distribution, and each pair has a half probability of being connected to form a virtual link. Besides, the demands of VNR for node and link resources are uniformly distributed from 0 to 50. 

All learning-based models are implemented with PyTorch.
% \cite{dl-pytorch}
 The trainable weights are initialized with the Xavier normal method 
% \cite{dl-xavier} 
and updated with the Adam optimizer \cite{dl-adam}. Specifically, in the testing phase, the upper-level agent greedily chooses actions with the highest probability, while the lower-level agent develops multiple solutions using the beam search strategy. Then, the physical network will allocate resources utilizing the feasible solution with the least cost. Other parameter settings are divided into three groups, neural network, model training, and intrinsic reward, shown in Table \ref{table-params}. The hardware we utilize to train models is a server of Ubuntu 18.04.2 LTS with 48 Intel(R) Xeon(R) Silver 4214 CPU @ 2.20GHz and 1 Nvidia GeForce RTX 3090 GPU.

\subsubsection{Compared Baselines} 
To validate the effectiveness of the proposed HRL-ACRA, we use the $LA\_Rev$ and $AC\_Ratio$ defined in Section \ref{preliminaries} as main evaluation indicators and compare it with six existing algorithms, which cover three classical heuristic-based algorithms, including \textbf{GRC} \cite{vne-grc}, \textbf{NRM} \cite{vne-nrm} and \textbf{PL} \cite{vne-tpds-dc}, and four learning-based algorithms, including \textbf{MCTS} \cite{vne-mcts}, \textbf{A3C-GCN} \cite{vne-a3c-gcn}, \textbf{REINFORCE-CNN} \cite{vne-cnn-double-layer} and \textbf{GAE-BFS} \cite{vne-gae-bfs}. The hyper-parameters of baseline algorithms are set to be the same as those mentioned in their original papers.

\subsection{Result Analysis}

The realistic network environment is usually complex and changeable. We simulate various network environments by adjusting some environment settings to verify the performance of these algorithms in dealing with real scenarios.

\subsubsection{\revise{Convergence Analysis}}
\begin{figure}[t] 
\vspace{-0.2cm}
 \setlength{\belowcaptionskip}{-0.1cm}
	\centering
	\includegraphics[width=.47\textwidth]{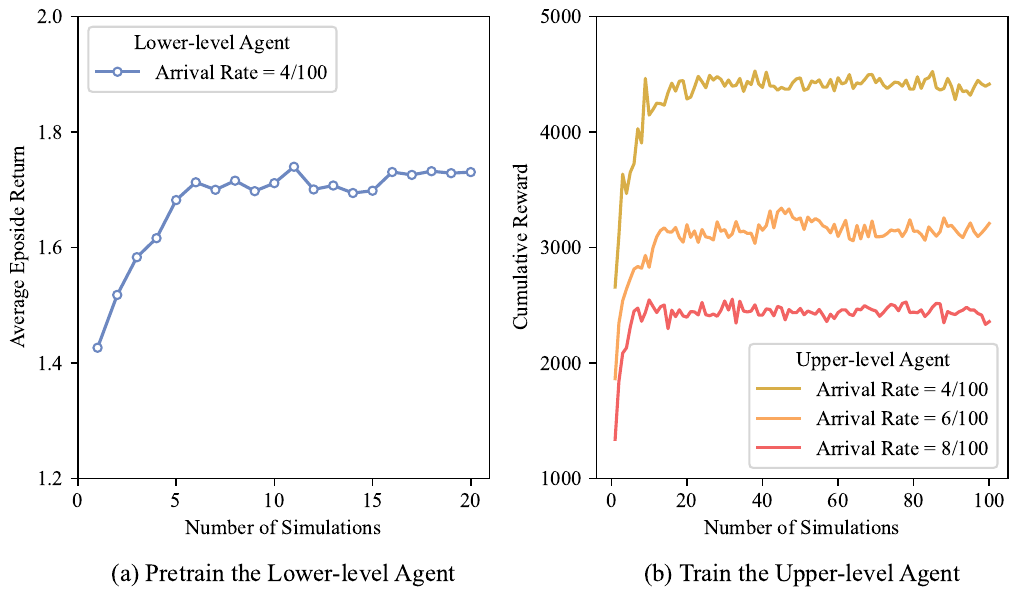} 
%	\vspace{-0.6cm}
	\caption{\revise{Learning curves of the lower-level and upper-level agent.}}
\label{fig-learning-curves}
\end{figure}

\revise{
The learning curves on pretraining the lower-level agent and jointly training two agents in different settings of arrival rate are illustrated in Figure \ref{fig-learning-curves}. \revisetwo{In each simulation, the lower-level agent focusing on VNR resource allocation and the upper-level agent having an infinite horizon, experience 1,000 episodes and 1,000 timesteps, respectively, to optimize their policies.}
We can observe that the lower-level agent achieves fast convergence at about the sixth episode in the pretraining process. 
\revisetwo{Due to the infinite horizon of the upper-level agent, we opt to assess the learning progress by utilizing the cumulative reward of one simulation, namely, the total reward accumulated over 1,000 timesteps.}
In the training process of three arrival rate settings, the upper-level agent also achieves fast convergences. 
HRL-ACRA can converge after an acceptable amount of training, which  manifests our reward design's effectiveness in guiding policy optimization.
}

\begin{table}[t]
\small
\centering
\caption{Average Running Time$^*$ in Node Size Tests}
\vspace{-0.2cm}
\begin{tabular}{c|ccc}
\hline
\multirow{2}{*}{Method}  & \multicolumn{3}{c}{Node Size}                                                  \\ \cline{2-4} 
                         & \multicolumn{1}{c}{2-10} & \multicolumn{1}{c}{2-15} & \multicolumn{1}{c}{2-20} \\ \hline
GRC   &  $128.01$   & $141.56$  &  $155.35$   \\
NRM   &   $95.54$   &  $\textbf{100.83}$    &   $\textbf{106.33}$   \\
\revise{PL}   &   \revise{$190.26$}  &  \revise{$238.06$}    &   \revise{$352.61$}   \\
MCTS    &  $953.38$  &  $2047.33$  &  $3244.17$  \\
A3C-GCN   &  $169.91$   &   $184.03$   & $285.41$  \\
REINFORCE-CNN   &  $133.41$   &   $180.10$   & $233.84$  \\
GAE-BFS   & $\textbf{98.41}$  &   $140.29$   &   $175.18$   \\
\textbf{HRL-ACRA (Ours)} &  $161.27$    &  $173.82$   &  $205.37$    \\ \hline
\end{tabular}
\label{table-running-time}
\begin{threeparttable}
 \begin{tablenotes}
        \footnotesize
        \item \revise{\hspace{0.2cm} $^*$Average of 10 experiments results (in seconds)}
  \end{tablenotes}
\end{threeparttable}

\end{table}

\begin{table*}[t]
\small
\centering
\caption{Arrival Rate Tests}
\vspace{-0.2cm}
\begin{tabular}{c|cc|cc|cc}
 \hline
\multicolumn{1}{c|}{\multirow{2}{*}{{Method}}} & \multicolumn{2}{c|}{Arrival Rate (4/100)}  & \multicolumn{2}{c|}{Arrival Rate (6/100)}  & \multicolumn{2}{c}{Arrival Rate (8/100)}  \\ \cline{2-7} 
\multicolumn{1}{c|}{}  & \multicolumn{1}{c}{AC\_Ratio} & \multicolumn{1}{c|}{LA\_Rev} & \multicolumn{1}{c}{AC\_Ratio} & \multicolumn{1}{c|}{LA\_Rev} & \multicolumn{1}{c}{AC\_Ratio} & \multicolumn{1}{c}{LA\_Rev} \\ \hline
GRC    & $81.96 \pm 2.64$ & $22.58 \pm 0.61$ & $66.48 \pm 3.04$ & $25.39 \pm 2.89$  & $58.63 \pm 2.71$ & $27.39 \pm 3.03$  \\ 
NRM    & $77.55 \pm 2.05$ & $20.71 \pm 0.81$ & $64.48 \pm 2.90$ & $23.91 \pm 2.53$ & $55.90 \pm 2.01$ & $25.48 \pm 2.82$  \\
\revise{PL} & \revise{$86.69 \pm 2.02$} & \revise{$23.63 \pm 0.62$} & \revise{$72.64 \pm 2.27$} & \revise{$29.43 \pm 3.00$} & \revise{$63.43 \pm 2.38$} & \revise{$31.60 \pm 3.50$} \\
MCTS   & $81.92 \pm 2.14$ & $21.12 \pm 0.68$ &  $71.14 \pm 2.52$ & $25.68 \pm 2.76$ & $62.77 \pm 2.24$ & $27.90 \pm 3.24$  \\
A3C-GCN   &  $87.70 \pm 1.61$ & $24.93 \pm 0.64$ & $73.91 \pm 1.92$ & $29.49 \pm 2.58$ & $63.21 \pm 2.53$ & $30.72 \pm 2.88$ \\
REINFORCE-CNN & $87.49 \pm 2.63$ & $24.79 \pm 0.83$ & $72.61 \pm 2.34$ & $28.43 \pm 2.72$  & $63.68 \pm 2.91$ & $30.71 \pm 2.78$ \\
GAE-BFS   & $88.71 \pm 1.64$ & $25.11 \pm 0.75$ & $76.00 \pm 2.65$ & $29.57 \pm 3.31$ &  $66.20 \pm 2.51$ & $31.07 \pm 3.05$ \\
\textbf{HRL-ACRA (Ours)}  & $\textbf{91.35} \pm \textbf{2.09}$  & $\textbf{26.18} \pm \textbf{0.73}$  & $\textbf{78.84} \pm \textbf{2.62}$  & $\textbf{31.52} \pm \textbf{2.77}$  &   $\textbf{69.73} \pm \textbf{2.63}$  &$\textbf{34.76} \pm \textbf{2.83}$ \\
\hline
\end{tabular}
\label{table-arrival-rate}
\end{table*}

\begin{table*}[t]
\small
\centering
\caption{Resource Request Tests}
\vspace{-0.2cm}
\begin{tabular}{c|cc|cc|cc}
 \hline
\multicolumn{1}{c|}{\multirow{2}{*}{{Method}}} & \multicolumn{2}{c|}{{Resource Request} (0-50)}  & \multicolumn{2}{c|}{{Resource Request} (0-60)}  & \multicolumn{2}{c}{{Resource Request} (0-70)}  \\ \cline{2-7} 
\multicolumn{1}{c|}{}  & \multicolumn{1}{c}{AC\_Ratio} & \multicolumn{1}{c|}{LA\_Rev} & \multicolumn{1}{c}{AC\_Ratio} & \multicolumn{1}{c|}{LA\_Rev} & \multicolumn{1}{c}{AC\_Ratio} & \multicolumn{1}{c}{LA\_Rev} \\ \hline
GRC   & $81.96 \pm 2.64$ & $22.58 \pm 0.61$ & $70.67 \pm 1.73$ & $20.99 \pm 0.87$ &  $61.20 \pm 2.36$ & $19.32 \pm 0.76$  \\ 
NRM & $77.55 \pm 2.05$ & $20.71 \pm 0.81$ & $66.58 \pm 2.61$ & $19.23 \pm 0.64$ & $58.84 \pm 2.10$ & $17.90 \pm 0.69$  \\
\revise{PL} & \revise{$86.69 \pm 2.02$} & \revise{$23.63 \pm 0.62$} & \revise{$76.20 \pm 2.11$} & \revise{$23.75 \pm 0.77$} & \revise{$66.27 \pm 2.45$} & \revise{$22.01 \pm 0.86$} \\
MCTS   & $81.92 \pm 2.14$ & $21.12 \pm 0.68$ & $72.80 \pm 1.92$ & $19.91 \pm 0.61$  & $64.25 \pm 1.87$ & $18.23 \pm 0.35$  \\
A3C-GCN   & $87.70 \pm 1.61$ & $24.93 \pm 0.64$ & $77.19 \pm 2.20$ & $23.74 \pm 0.76$ & $67.58 \pm 2.35$ & $21.95 \pm 0.78$ \\
REINFORCE-CNN & $87.49 \pm 2.63$ & $24.79 \pm 0.83$ & $75.68 \pm 1.98$ & $22.90 \pm 0.64$ & $66.26 \pm 2.42$ & $21.26 \pm 0.67$ \\
GAE-BFS   & $88.71 \pm 1.64$ & $25.11 \pm 0.75$ & $78.65 \pm 1.60$ & $23.73 \pm 0.78$ & $69.26 \pm 1.83$ & $21.83 \pm 0.58$ \\
\textbf{HRL-ACRA (Ours)}  & $\textbf{91.35} \pm \textbf{2.09}$  & $\textbf{26.18} \pm \textbf{0.73}$  & $\textbf{82.38} \pm \textbf{2.01}$  & $\textbf{25.73} \pm \textbf{0.74}$  & $\textbf{73.65} \pm \textbf{3.28}$  &$\textbf{25.62} \pm \textbf{0.73}$ \\
\hline
\end{tabular}
\label{table-resource-request}
\end{table*}

\begin{table*}[t]
\centering
\small
\caption{Node Size Tests}
\vspace{-0.2cm}
\begin{tabular}{c|cc|cc|cc}
 \hline
\multicolumn{1}{c|}{\multirow{2}{*}{{Method}}} & \multicolumn{2}{c|}{{Node Size} (2-10)}  & \multicolumn{2}{c|}{{Node Size} (2-15)}  & \multicolumn{2}{c}{{Node Size} (2-20)}  \\ \cline{2-7} 
\multicolumn{1}{c|}{}  & \multicolumn{1}{c}{AC\_Ratio} & \multicolumn{1}{c|}{LA\_Rev} & \multicolumn{1}{c}{AC\_Ratio} & \multicolumn{1}{c|}{LA\_Rev} & \multicolumn{1}{c}{AC\_Ratio} & \multicolumn{1}{c}{LA\_Rev} \\ \hline
GRC   & $81.96 \pm 2.64$ & $22.58 \pm 0.61$ & $60.42 \pm 2.38$ & $22.08 \pm 0.94$ & $48.40 \pm 1.57$ & $20.87 \pm 0.69$ \\ 
NRM & $77.55 \pm 2.05$ & $20.71 \pm 0.81$ & $57.34 \pm 2.90$ & $20.02 \pm 1.10$ & $46.25 \pm 1.48$ & $18.79 \pm 0.51$ \\
\revise{PL} & \revise{$86.69 \pm 2.02$} & \revise{$23.63 \pm 0.62$} & \revise{$65.20 \pm 2.27$} & \revise{$25.20 \pm 1.25$} & \revise{$52.99 \pm 1.10$} & \revise{$23.78 \pm 0.48$} \\
MCTS   & $81.92 \pm 2.14$ & $21.12 \pm 0.68$ & $60.15 \pm 1.81$ & $18.46 \pm 0.61$	 & $47.47 \pm 1.50$ & $16.00 \pm 0.50$ \\
A3C-GCN   & $87.70 \pm 1.61$ & $24.93 \pm 0.64$ & $64.83 \pm 2.17$ & $23.87 \pm 0.87$ & $51.79 \pm 1.75$ & $22.01 \pm 0.42$ \\
REINFORCE-CNN & $87.49 \pm 2.63$ & $24.79 \pm 0.83$ & $65.40 \pm 2.77$ & $24.36 \pm 1.03$ & $52.73 \pm 1.66$ & $22.89 \pm 0.85$ \\
GAE-BFS   & $88.71 \pm 1.64$ & $25.11 \pm 0.75$ & $67.79 \pm 2.10$ & $24.47 \pm 1.01$ & $54.00 \pm 1.54$ & $22.20 \pm 0.63$	 \\
\textbf{HRL-ACRA (Ours)}  & $\textbf{91.35} \pm \textbf{2.09}$  & $\textbf{26.18} \pm \textbf{0.73}$  & $\textbf{69.51} \pm \textbf{2.93 }$  & $\textbf{26.63} \pm \textbf{0.94}$  & $\textbf{55.34} \pm \textbf{1.58}$  &$\textbf{25.58} \pm \textbf{0.84}$ \\
\hline
\end{tabular}
\label{table-node-size}
\end{table*}

\begin{figure*}[t] 
\vspace{-0.2cm}
 \setlength{\belowcaptionskip}{-0.1cm}
	\centering
	\includegraphics[width=.95\textwidth]{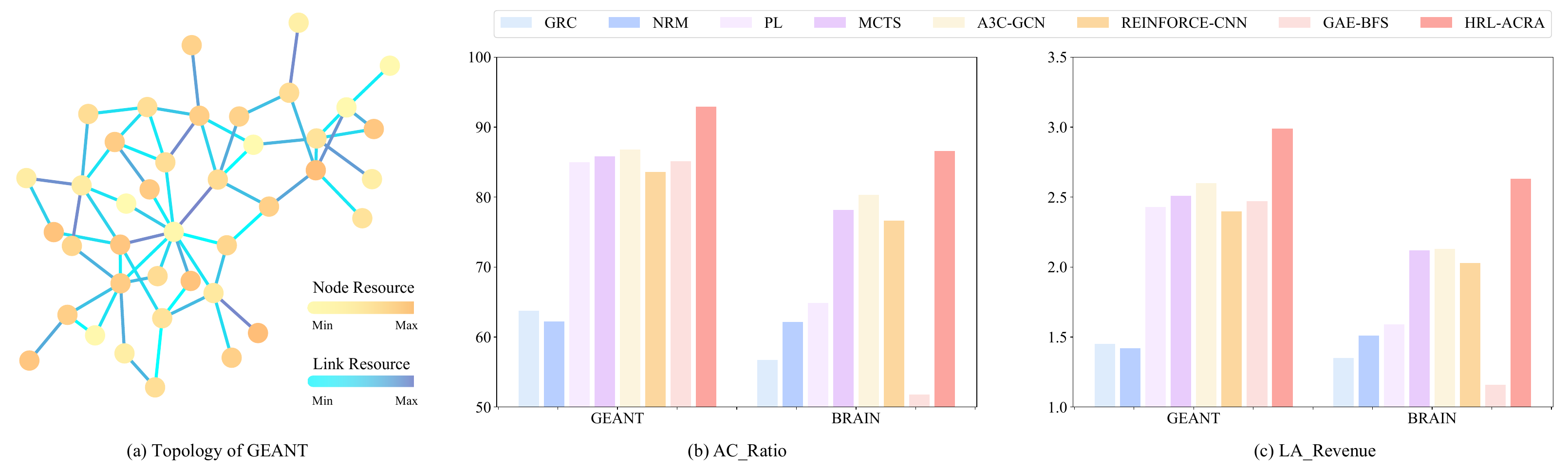} 
	% \vspace{-0.6cm}
	\caption{Validation on real network topologies. (a) shows the topology of GEANT, where node and edge colors indicate different computing and bandwidth resource capacities, respectively. (b) and (c) presents experimental results on the acceptance ratio and long-term average revenue.}
\label{fig-real-network}
\end{figure*}

\subsubsection{Arrival Rate Tests} 
In different periods, network systems often confront various service request traffic. To evaluate the ability of VNE algorithms to handle various frequencies of VNRs, we run these algorithms in three frequent arrival scenarios by setting the arrival rate to 0.04, 0.06, and 0.08, respectively. The results are presented in Table \ref{table-arrival-rate}.

We can see that as the arrival rate of VNRs increases, acceptance rates of all algorithms decrease because more VNRs compete for the resources of the physical network in the same period. 
\revise{When the arrival rate is set to 0.08, our proposed HRL-ACRA increases the $AC\_Ratio$ by about 18.93\%, 10.31\%, and 5.33\%  compared to the GRC, A3C-GCN, and GAE-BFS, respectively. Similarly, our proposed method achieves 26.91\%, 13.15\%, and 11.88\% improvement in $LA\_Rev$, compared to GRC, A3C-GCN, and GAE-BFS.}
Thanks to the upper-level agent for admission control, our proposed method can promote long-term benefits by rejecting several inappropriate VNRs in the early stage, alleviating the competition for physical resources.

\subsubsection{Resource Request Tests} 
Resource-intensive VNRs tend to require more resources while others are with less resource demands.
To reflect the effectiveness of VNE algorithms that embeds VNRs with various resource demands, we imitate three scenarios by adjusting the upper bound of resource requests to 50, 60, and 70.

Table \ref{table-resource-request} illustrates that HRL-ACRA outperforms all the baselines in terms of the $AC\_Ratio$ and $LA\_Rev$ in this group experiment.
With the increment of the upper bound of resource requests, the acceptance rates of all algorithms are declining, while the average long-term revenues are relatively stable.
Our customized GNN takes full advantage of node and link features, capable of allocating resources reasonably according to the real-time situation of the physical network. The multi-objective intrinsic reward could also guide the low-level agent to achieve resource load balancing and maximize the revenue-cost ratio, which reserves more resources for subsequently arriving VNRs.

\subsubsection{Node Size Tests} 
When the scale of the services is different, the number of virtual nodes of VNRs changes accordingly. For example, enterprise-level services are often extensive, while ordinary user services require few nodes.
To measure the algorithm's generalizability, we mimic large, medium, and microservices by changing the node size maximum to 10, 15, and 20, respectively. 

Table \ref{table-node-size} reports the experiment results in these three settings. From the results shown in Table \ref{table-node-size}, we observe that our HRL-ACRA consistently achieves the highest $LA\_Rev$ and $AC\_Ratio$ compared to other algorithms. 
Iteratively constructing resource allocation solutions through the seq2seq model significantly reduces the action space, speeding up training and improving generalization.
Moreover, the deep feature-aware GNN encoder and positional encoder extract sufficient information on VNR's topology structure and decision order, providing more effective node representations to generate better embedded actions. It demonstrates the adaptability of  HRL-ACRA to large-sized VNRs.

\subsubsection{Running Time Tests}
To verify the response speed of VNE algorithms, we use the average running time for tackling 1000 VNRs as the evaluation metric shown in Table \ref{table-running-time} with various node sizes. Except for MCTS based on a massive search, our proposed method and other algorithms can allocate resources to VNRs in an acceptable time. Our proposed algorithm can process the VNR arriving online in real-time, attributing to the fast inference ability of deep neural networks. 
% Additionally, our proposed method with a more complex neural network structure is faster than A3C-GCN method, which indicates that the admission control mechanism saves the inference time in the resource allocation stage by early rejecting some non-embeddable VNRs.
\revise{Additionally, the running time of HRL-ACRA demonstrates a minimal increase with the increase of the node size.
This is due to the fact that, as the node size increases, the failure probability of resource allocation also rises. 
However, HRL-ACRA is able to avoid some resource allocation processes by allowing the upper-level agent to reject non-embeddable VNRs, thus reducing overall time consumption.}

\subsection{Validation on Real Network Topologies}

Similar to previous works \cite{vne-a3c-gcn},\cite{vne-tsc-energy}, we conduct additional experiments on real network topologies, GEANT and BRAIN, to verify the feasibility of our proposed algorithm when applied in realistic network environments.
\revise{GEANT is the academic research network interconnecting Europe's national research and education networks, with 40 nodes and 64 edges, whose topology is shown in Figure \ref{fig-real-network}(a). 
BRAIN, consisting of 161 nodes and 166 edges, is the high-speed data network for scientific and cultural institutions in Berlin.
Compared with GEANT with a density of 0.0821, BRAIN with a density of 0.0129 is larger but sparser.
Except that the node and link resource demand of VNRs are uniformly distributed from 0 to 5, other environment simulation parameters are the same as those discussed in Section \ref{experimental-setup}.
As shown in Figures \ref{fig-real-network}(b) and \ref{fig-real-network}(c), the performance of GAE-BFS drops rapidly when applied to BRAIN, because GAE-BFS relies seriously on backtracking after failure so that it struggles to effectively explore the search space, lacking adaptation to large-scale and sparse topologies.
However, HRL-ACRA still outperforms these baselines on both the acceptance ratio and long-term average revenue at both two topologies, demonstrating the robustness of applying our proposed algorithm in various real network topologies.}

\section{Conclusion}
\label{conclusion}
\revisethree{In this paper, we have introduced a novel optimization perspective and proposed a hierarchical RL framework to tackle the VNE problem. } Our framework consists of two components, i.e., an upper-level agent and a lower-level agent. The upper-level agent aims to decide whether to admit the arriving VNR or not, while the lower-level tries to generate high-quality solutions of resource allocation for admitted VNRs.
To address the infinite horizon problem of the upper-level agent, we have adopted the average revenue method to achieve the tradeoff of the current reward and future return.
We also have designed a customized multi-objective intrinsic reward with multiple local indicators to alleviate the sparse reward issue of the lower-level agent.
Moreover, leveraging the deep feature-aware GNN and the GRU-based seq2seq model, HRL-ACRA has been able to capture the temporal dependence and topological structure of VNRs and physical network. \revisethree{Extensive experimental results have demonstrated that our HRL-ACRA outperforms other SOTA approaches, particularly achieving significant improvement in more practical resource-limited scenarios.}

In our further work, we intend to integrate more constraints (e.g., path latency and failure guarantee) into our proposed framework, which enhances the algorithm to handle more real-world network environments. 
We also plan to solve VNE problems with combinatorial objectives to meet the diverse indicators of Internet providers, such as joint consideration of electricity costs.

% \clearpage
\bibliography{main}
\bibliographystyle{IEEEtran}

\clearpage

\appendices

\section{Baseline Descriptions}

In this section, we provide detailed descriptions of baseline algorithms.

\begin{itemize}[leftmargin=*]
  \item \textbf{GRC} \cite{vne-grc} is a heuristic-based algorithm that applies global resource management to rank the importance of nodes considering resource features and topological attributes. Then, it maps virtual nodes onto physical nodes with a greedy matching strategy. An admission control strategy based on revenue-to-cost also is utilized to improve the long-term benefits.
  \item \textbf{NRM} \cite{vne-nrm} is a heuristic-based algorithm that develops a measurement considering multi-type resources to evaluate the resource sufficiency of the physical nodes. It adopts this measurement to choose the physical node with the most adequate resources to execute the node mapping and utilizes the shortest path algorithm to perform the link mapping.

  \item \revise{ \textbf{PL} \cite{vne-tpds-dc} is a heuristic-based algorithm that develops node proximity sensing and path comprehensive evaluation to enhance the node and link mapping correlation. The result of the iterative selection of nodes will also affect the path assessment simultaneously to realize the coordination of node and link mapping.}

  \item \textbf{MCTS} \cite{vne-mcts} is an RL-based algorithm based on MCTS to search the action space. It first calculates each physical node's upper confidence bound (UBC) value to estimate the long-term benefits. Subsequently, it uses a greedy strategy for node mapping based on UBC values and then performs link mapping.
  
  \item \textbf{A3C-GCN} \cite{vne-a3c-gcn} is an RL-based algorithm that employs GCN to extract information of states and uses A3C to accelerate the training efficiency. 
  However, since the original GCN can not exploit link features, this algorithm roughly aggregates link resources into node features.
 
   \item \textbf{REINFORCE-CNN} \cite{vne-cnn-double-layer} is an RL-based algorithm that uses CNN to extract the features of physical nodes. In addition to the resource availability of physical nodes, it also leverages some graph theory features to inject information on topology structure, such as degree centrality, closeness centrality, etc.
 
  \item \textbf{GAE-BFS} \cite{vne-gae-bfs} is an uSL-based algorithm that adopts GAE to learn latent embedding of physical nodes and uses them to cluster physical nodes. Then, based on random sampling within each cluster, it selects several initial nodes to execute BFS methods and finally chose the solution with the least cost.

\end{itemize}

\section{Additional Experiments}

In this section, we present additional experimental results and analyses to further validate the effectiveness of our proposed algorithm. Specifically, we conduct ablation studies and study the impact of pricing strategies and the sensitivity of hyperparameters.

\subsection{Ablation Studies}

\begin{figure}[t] 
\vspace{-0.2cm}
 \setlength{\belowcaptionskip}{-0.1cm}
	\centering
	\includegraphics[width=.47\textwidth]{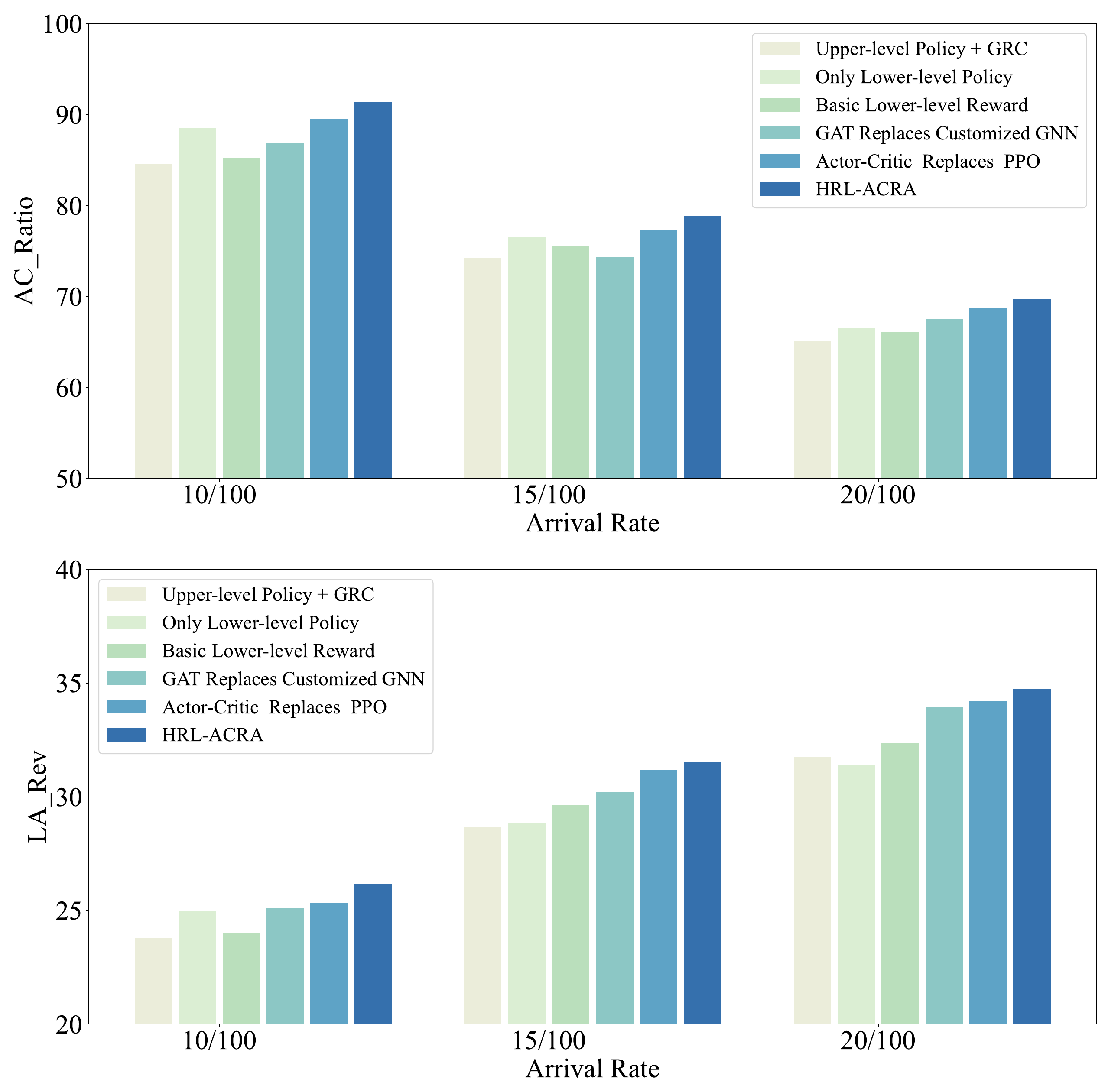} 
%	\vspace{-0.6cm}
	\caption{Evaluation results on ablation studies. The acceptance ratio and long-term average revenue in various arrival rate settings.} 
\label{fig-ablation-studies}
\end{figure}

To study the effectiveness of each component of our HRL-ACRA, we conducted ablation studies by training several variants of our model, including
\begin{itemize}
    \item Upper-level Policy + GRC: This method replaces our lower-level policy with GRC.
    \item Only Lower-level Policy: This method abandons the upper-level agent and only utilizes our lower-level policy.
    \item Basic Lower-level Reward: This method uses a  basic lower-level reward function, which is as follows
    \begin{equation}
        r = 
        \begin{cases}
        \frac{\text{Rev}({G}^v)}{\text{Cost}({G}^v)}, & \text{if} \ G^v \ \text{is embedded successfully}, \\
        0 & \text{otherwise},
    \end{cases}
    \end{equation}
    \item \revisetwo{GAT Replaces Our Customized GNN: This method uses the original GAT to extract features from states.}
    \item \revisetwo{Actor-Critic Replaces PPO: This method uses the vanilla Actor-Critic instead of PPO to optimize the policies.}
\end{itemize}

Figure \ref{fig-ablation-studies} depicts that HRL-ACRA surpasses these variants on $AC\_Ratio$ and $LA\_Rev$. 
Through comparisons between Upper-level Policy + GRC, Only Lower-level Policy, and HRL-ACRA, it can be seen that learning a joint policy of admission control and resource allocation can effectively improve the performance of VNE. The intrinsic reward designed in HRL-ACRA also encourages the lower-level agent to conduct better exploration, as shown in the comparison between Basic Lower-level Reward and HRL-ACRA. 
\revisetwo{Compared to using GAT as a feature extractor, HRL-ACRA equipped with our customized GNN enables the agents to perceive more precise information about states and achieve better performance. Moreover, using PPO as the training algorithm is able to achieve more efficient exploration compared to using the vanilla actor-critic.} These results demonstrate that each component is critical to making our approach powerful.

\subsection{The Impact of Pricing Strategy}

In practical scenarios, internet providers adopt different pricing strategies to maximize their profits, based on their specific situations. Regulating the value of the start price weight ($w_a$) and the service time charge weight ($w_b$) can control the percentage of initialization-related costs (e.g., server configuration) and time-related costs (e.g., electricity fees). To investigate the effect of these two weights on performance, we conducted additional experiments comparing different service charge mechanisms. We first fixed the weight of the starting price ($w_a$) to 1 and then simulated different profit models by changing the service time charge weight ($w_b$). Intuitively, a smaller $w_b$ value indicates that the starting price accounts for the majority of the revenue, while a larger $w_b$ value indicates that the service time charge has a greater impact on profitability.
In these experiments, we set the arrival rate to 0.08, while other settings remained the same as the default settings described in Section 5.1.1.

\begin{figure}[h] 
\vspace{-0.2cm}
 \setlength{\belowcaptionskip}{-0.1cm}
	\centering
	\includegraphics[width=.48\textwidth]{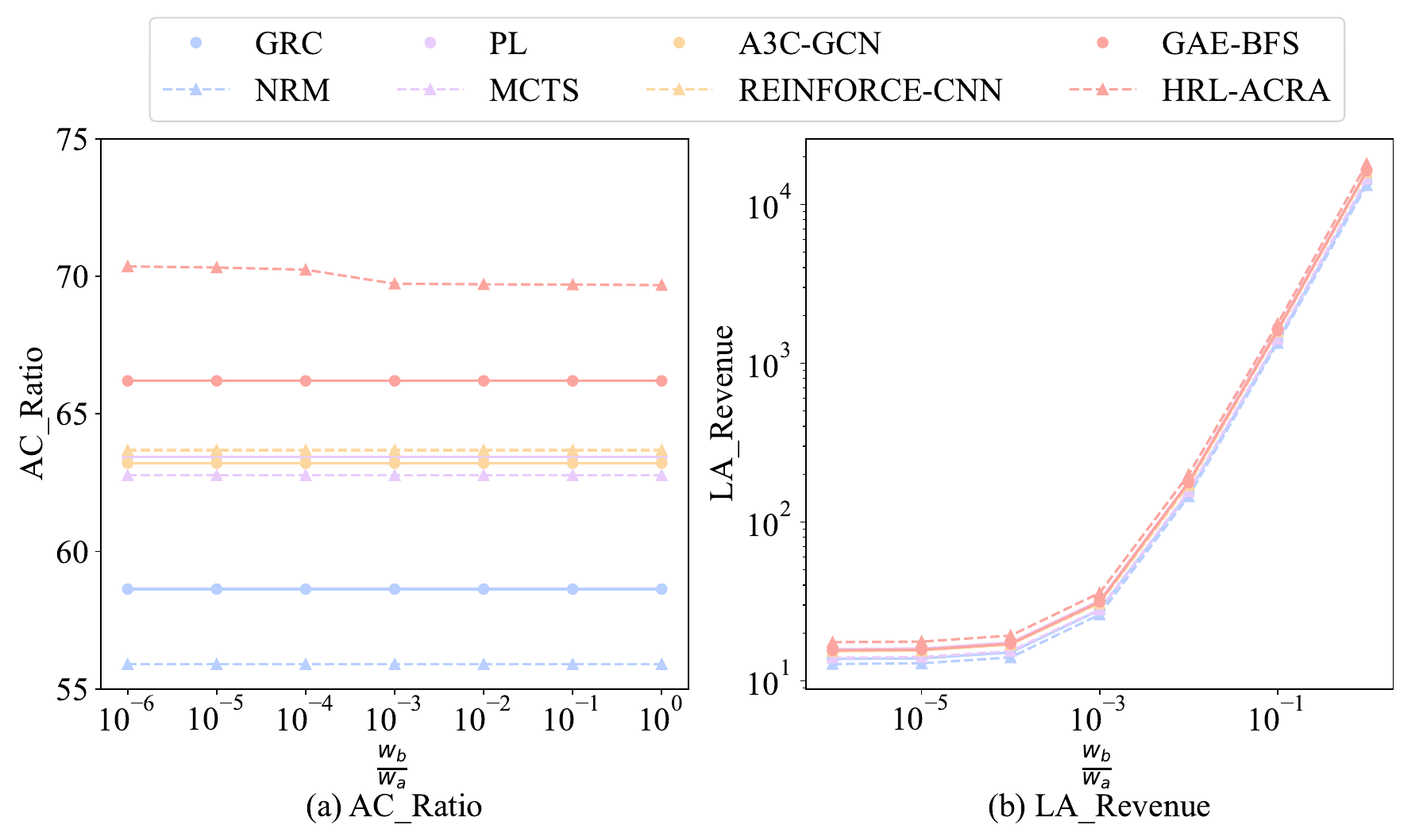} 
	\caption{Evaluation results on ablation studies. The acceptance ratio and long-term average revenue in various arrival rate settings.} 
\label{fig-charge-metrics}
\end{figure}

\begin{figure}[h]
\vspace{-0.2cm}
 \setlength{\belowcaptionskip}{-0.1cm}
	\centering
	\includegraphics[width=.48\textwidth]{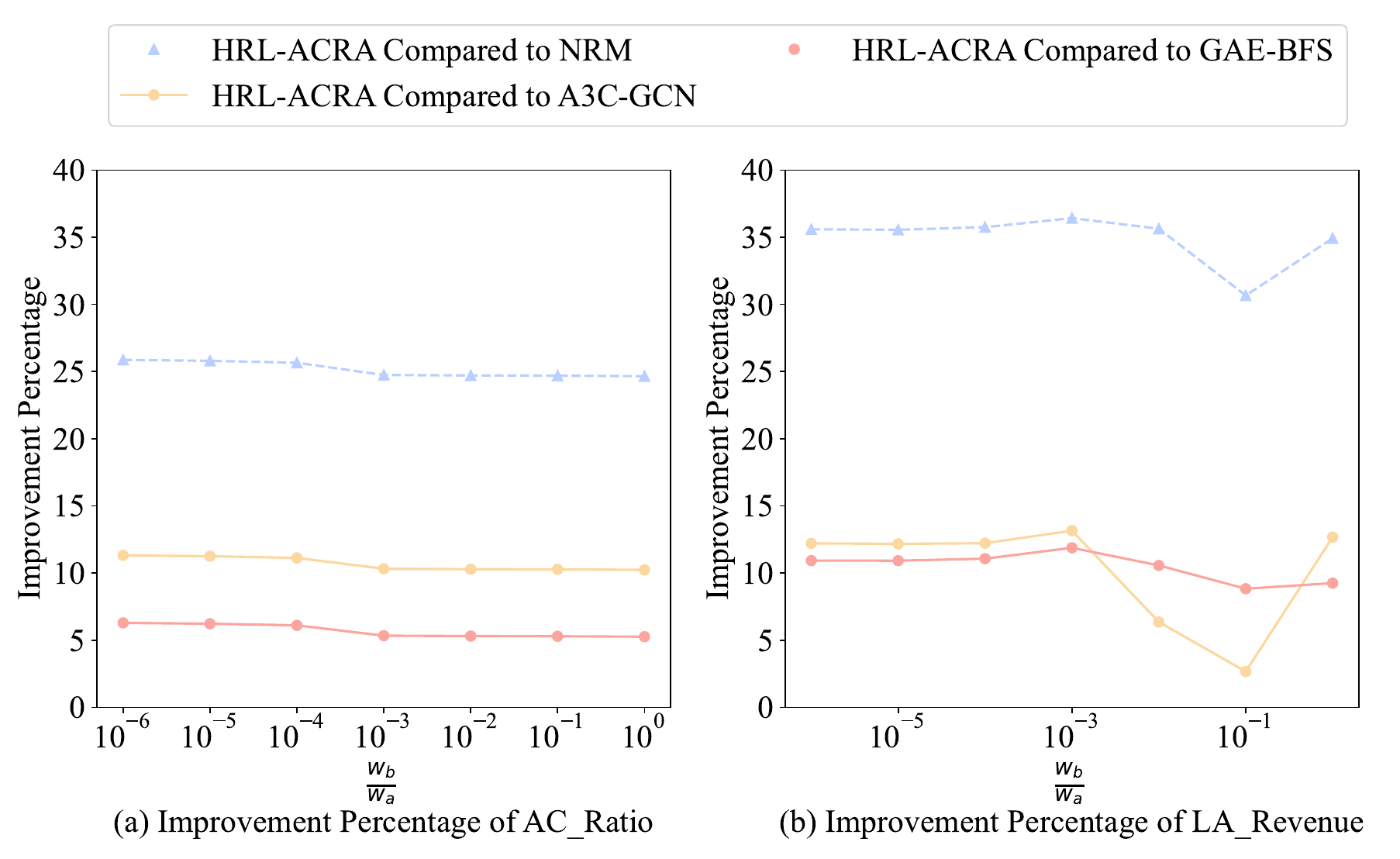} 
%	\vspace{-0.6cm}
	\caption{Evaluation results on ablation studies. The acceptance ratio and long-term average revenue in various arrival rate settings.} 
\label{fig-charge-improvement}
\end{figure}

From Figure \ref{fig-charge-metrics} (a) and Figure \ref{fig-charge-improvement} (a), we can observe that as $w_b$ increases, the acceptance rate of HRL-ACRA sightly decreases and other algorithms' acceptance rate remains unchanged, which mainly results from the adaptive admission control mechanism of HRL-ACRA's upper-level agent. Other algorithms lack the ability to directly optimize long-term average revenue, so changes in profit weights will not affect their performance on acceptance rate. Besides, With the increase of $w_b$, the service time charge become to play a more important role in profitability. Therefore, HRL-ACRA prefers to accept VNRs with longer lifetimes to reduce the fragmentation of resources, leading to a slight drop in HRL-ACRA's acceptance rate. From Figure \ref{fig-charge-metrics} (b) and Figure \ref{fig-charge-improvement} (b), we can see that all algorithms' long-term average revenue raise sharply, with the increase of $w_b$, which results from the proportion of service time charge is getting higher. Additionally, the changing trend of the improvement percentage of HRL-ACRA compared to other algorithms is not obvious, indicating that the weight change has little impact on the performance of the long-term average revenue. Overall, changes in profit weights have a weak effect on the acceptance rate of HRL-ACRA, while the effect on the long-term average return of HRL-ACRA is not significant.

\subsection{The Sensitivity of Hyper Parameters}

In regards to the parameters of neural networks and reinforcement learning, there is ample prior research providing guidance on their appropriate values. These parameters have a relatively minor impact on the performance of the neural network when set to reasonable values. As a result, we have determined their values using general settings that are widely used in related research. In this study, our main focus is to evaluate the effect of the reward weights $w_1$ and $w_2$ on the performance of HRL-ACRA, which are special parameters proposed in our work.

\begin{figure}[h]
\vspace{-0.2cm}
 \setlength{\belowcaptionskip}{-0.1cm}
	\centering
	\includegraphics[width=.48\textwidth]{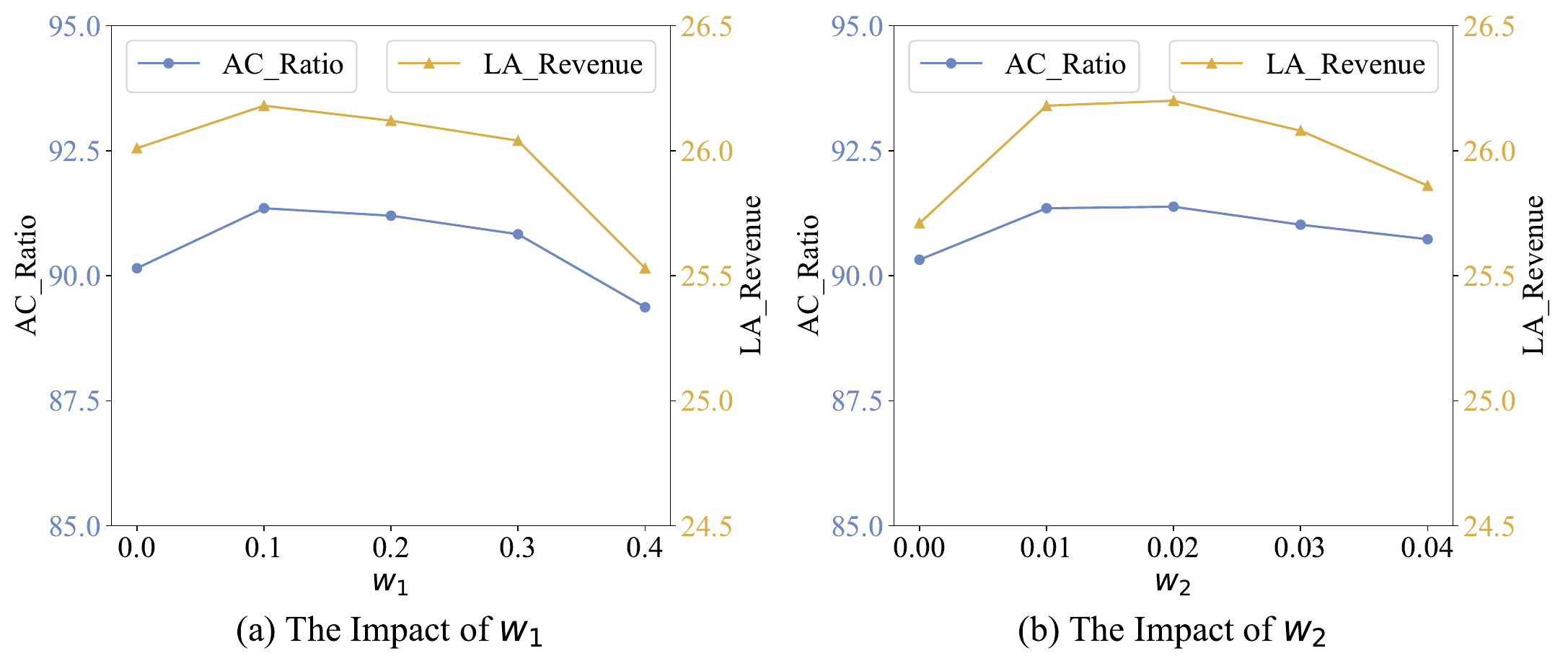} 
%	\vspace{-0.6cm}
	\caption{The impact of $w_1$ and $w_2$ on AC\_Ratio and LA\_Revenue.} 
\label{fig-parameter-sensitivity}
\end{figure}

First, we separately evaluate the effect of $w_1$ on the performance of HRL-ACRA by fixing $w_2 = 0.01$.
The results are shown in Figure \ref{fig-parameter-sensitivity}(a). By incorporating a penalty term to penalize inappropriate admission of VNRs that do not have infeasible solutions, the upper-level agent can more proactively reject VNRs to enhance the long-term benefits. However, a high value of the penalty parameter ($w_2$) can lead to a misguided rejection of many VNRs early, causing a reduction in the acceptance rate and long-term average benefits.
Furthermore, we can observe that the performance of HRL-ACRA is good enough when $w_1$ ranges from 0.1 to 0.2.
Second, we separately evaluate the effect of $w_2$ on the performance of HRL-ACRA by fixing $w_1 = 0.1$. 
The results are shown in Figure \ref{fig-parameter-sensitivity}(b). A well-resource load balancing can facilitate the placement of virtual nodes, leading to an enhanced success rate of embedding VNRs that arrive subsequently. Incorporating this aspect as a constituent of the reward function can lead to an improvement in both the acceptance ratio and long-term average revenue. However, as the weight $w_2$ increases, it may override the primary optimization objective of the lower-level agent, i.e., maximizing the revenue-to-cost ratio, which may lead to a compromised quality of the VNR embedding solution and consequently, a deterioration in the performance of the HRL-ACRA. Moreover, it is noteworthy that the HRL-ACRA system delivers a satisfactory performance within the range of $w_1$ from 0.01 to 0.03.
The preceding experiments provide empirical evidence that the HRL-ACRA exhibits robustness, and the parameters are suitably selected within a reasonable range.

\end{document}